\documentclass{emulateapj}
\usepackage{lscape}
\input{epsf.sty}

\newcommand{\solar}{\ifmmode_{\mathord\odot}\else$_{\mathord\odot}$\fi} % sun
\newcommand{\dgr}{\ifmmode {^\circ}\else $^\circ $\fi}  % arc se
\newcommand{\arcs}{\ifmmode {'' }\else $'' $\fi}  % arc se
\newcommand{\arcm}{\ifmmode {' }\else $' $\fi}    % arc min
\newcommand{\mstar}{\ifmmode {M_{HI_\ast}}\else $M_{HI_\ast}$\fi}
\newcommand{\msolar}{\ifmmode {M_\odot} \else M$_\odot$\fi}
\newcommand{\zsun}{\ifmmode {Z_\odot} \else Z$_\odot$\fi}
\newcommand{\gapprox}{\ifmmode \buildrel > \over {_\sim} \else $\buildrel >\over {_\sim}$\fi}
\newcommand{\lapprox}{\ifmmode \buildrel < \over {_\sim} \else $\buildrel <\over {_\sim}$\fi}

\newcommand{\ha}{\ifmmode H\alpha \else H$\alpha$\fi}
\newcommand{\msunyr}{\ifmmode {M_\odot yr^{-1}} \else M$_\odot$ yr$^{-1}$\fi}
\newcommand{\darcs}{\ifmmode \buildrel {''} \over . \else \buildrel $''$ \over . \fi}
\newcommand{\Udiff}{\ifmmode U^{1-\alpha}_{min} - U^{1-\alpha}_{max} \else U$^{1-\alpha}_{min} -$ U$^{1-\alpha}_{max}$ \fi}

\begin{document}

\shortauthors{Rosenberg, Wu, Le Floc'h, Charamandris, Ashby, Houck, Salzer, \& Willner}

\title{Dust Properties and Star-Formation Rates in Star-Forming Dwarf Galaxies}

\author{J. L. Rosenberg\altaffilmark{1}}
\affil{Harvard-Smithsonian Center for Astrophysics, 60 Garden Street MS 65, 
Cambridge, MA 02138}
\email{jrosenb4@gmu.edu}
\author{Yanling Wu}
\affil{Astronomy Department, Cornell University, 610 Space Sciences Building,
Ithaca, NY 14853}
\author{Emeric Le Floc'h\altaffilmark{2,3}}
\affil{Steward Observatory, University of Arizona, 933 North Cherry Avenue,
Tucson, AZ 85721.; Chercheur Associ\'e, Observatoire de Paris, F-75014, Paris,
France}
\author{V. Charmandaris\altaffilmark{2,4}}
\affil{Department of Physics, University of Crete, P.O. Box 2208, GR-71003
Heraklion, Greece} 
\author{M. L. N. Ashby}
\affil{Harvard-Smithsonian Center for Astrophysics, 60 Garden Street MS 65, 
Cambridge, MA 02138}
\author{J. R. Houck}
\affil{Astronomy Department, Cornell University, 610 Space Sciences Building,
Ithaca, NY 14853}
\author{J. J. Salzer} 
\affil{Wesleyan University, Department of Astronomy, Middletown, CT 06459}
\author{S. P. Willner}
\affil{Harvard-Smithsonian Center for Astrophysics, 60 Garden Street MS 65, 
Cambridge, MA 02138}

\altaffiltext{1}{current address: George Mason University,
Dept. of Physics and Astronomy, Fairfax, VA 22030}
\altaffiltext{2}{Chercheur Associe,  Observatoire de Paris, F-75014, Paris, France}
\altaffiltext{3}{current address: Spitzer Fellow, Inst. for Astronomy, 
Univ. of Hawaii, 2680 Woodlawn Drive, Honolulu HI 96822}
\altaffiltext{4}{IESL/Foundation for Research and Technology - 
Hellas, GR-71110, Heraklion, Greece}

\begin{abstract}

We have used the {\it Spitzer Space Telescope} to study the dust properties of a 
sample of star-forming dwarf galaxies. The differences in the mid-infrared
spectral energy distributions for these galaxies which, in general, are low
metallicity systems, indicate differences in the physical properties, heating,
and/or distribution of the dust. Specifically, these galaxies have more hot dust
and/or very small grains and less PAH emission than either spiral or higher
luminosity starburst galaxies. As has been shown in previous studies, there is a 
gradual decrease in PAH emission as a function of metallicity. Because much of
the energy from star formation in galaxies is re-radiated in the mid-infrared, 
star-formation rate indicators based on both line and continuum measurements in 
this wavelength range are coming into more common usage. We show that the
variations in the interstellar medium properties of galaxies in our sample, as
measured in the mid-infrared, result in over an order of magnitude spread in the
computed star-formation rates. 

\end{abstract}


\section{INTRODUCTION}

Mid-infrared (MIR) observations probe the dusty interstellar medium and the
dust-enshrouded star formation in galaxies. The spectral energy distribution
in this region is shaped by this star formation, but can also be affected by the 
presence (or lack) of dust grains of varying size, by differences in
metallicity, and by radiation from an older population of stars \citep{li2002}.
One of the first studies of the MIR spectra of a large sample of galaxies
including the low metallicity system II Zw 40 was performed by
\citet{roche1991}. More
recent observations have been used to understand how these different properties 
affect a galaxy's spectral energy distribution. With the advent of the 
{\it Spitzer Space Telescope} and the {\it Infrared Space Observatory}, the 
facilities are now available to make deep observations in the MIR. These data 
are being used to study the interstellar medium and to measure the
star-formation rate (SFR) in galaxies both in the nearby and distant universe. 

Some detailed, broadband, MIR observations of low metallicity
galaxies have shown that these systems display a wide range of colors which
implies a wide range of interstellar medium properties \citep{rosenberg2006,
engelbracht2005,hogg2005}.
A more detailed examination of the dust properties of low metallicity galaxies
has been performed for some of the best known low metallicity galaxies using
infrared spectroscopy \citep{wu2007,wu2006,houck2004b, hunt2005,hunt2006,
madden2006}.
These observations have provided evidence that there is a relationship between 
how much polycyclic aromatic hydrocarbon (PAH) emission is detectable in these 
systems and their metallicity. In the two lowest metallicity galaxies known, 
I Zw 18 and SBS 0335-052, no PAH emission is detected even in deep 
observations \citep{wu2007,houck2004b, thuan1999}. Several
models for the decrease in PAH emission with metallicity have been proposed
including destruction of PAH molecules by the hard radiation field in low
metallicity systems because the dust opacity is low in these environments
\citep{galliano2005,madden2006}, destruction in the shocks created by
supernova blast waves \citep{ohalloran2006}, and an intrinsic lack of PAH 
molecules in young systems because the PAHs are produced in low-mass stars that
have not yet evolved \citep{dwek2005,galliano2007}.

Using the measured MIR properties of galaxies to determine their star formation 
rates requires an understanding of the properties that affect the heating of
the dust responsible for the MIR emission. At high redshift, in particular, the 
effect of metallicity on the MIR emission may become more significant if a
larger fraction of the
galaxy population consists of low metallicity systems. The relationships between 
MIR flux density and SFR may not be the same as those for the more luminous and 
metal-rich systems for which the relationships were determined (e.g., 
\citealp{rosenberg2006,wu2006,engelbracht2005}). Emission from very small grains 
and warmer dust is seen in low metallicity galaxies but much less emission from
PAH molecules is detected relative to spiral and higher luminosity starburst 
galaxies \citep{madden2005,hunt2005}. For example, in the Small Magellanic Cloud 
where the metallicity is low, the dust properties, absorption, and emissivity 
are significantly different from what is seen in the Milky Way, indicating that 
metal-rich galaxies are not a proxy for their lower-metallicity counterparts 
\citep{li2006}. 

The star-forming dwarf galaxies that are discussed here
have been studied previously in the {\it Spitzer}/IRAC bands and have been found
to have a wide range in mid-infrared colors \citep{rosenberg2006}. Here we look
in greater detail at the spectral energy distributions for these systems as well
as the relationship between dust properties and metallicity, and the
measurements of star-formation in the mid-infrared and ultraviolet. The data are
presented in \S 2. A detailed look at flux density ratios and the MIR SEDs of
these galaxies are examined in \S 3 to answer
the question, as best as the data will allow, do these galaxies exhibit PAH
emission? The measured SFRs determined from the optical, MIR, and ultraviolet 
(UV) data are discussed and compared in \S 4, and the results are summarized in
\S 5.

\section{DATA}

\subsection{Sample Selection and Optical Data}

The galaxies in this study are low luminosity star-forming galaxies 
selected from the KPNO International Spectroscopic Survey (KISS) in the Bo\"otes
field. These systems have been observed both as a part of the KISS survey and by
several other surveys at optical, infrared, and ultraviolet wavelengths
including the NOAO Deep Wide-Field Survey (NDWFS,
\citealp{jannuzi1999,jannuzi2007}), the {\it Spitzer}/IRAC Shallow
Survey \citep{eisenhardt2004}, and a shallow {\it Spitzer}/MIPS survey.
\citet{rosenberg2006} provides a discussion of
the {\it Spitzer}/IRAC Shallow Survey data for these objects. 

KISS is a modern objective-prism survey. It combines the methodology of
many of the classic wide-field color- and line-selected surveys \citetext{e.g.,
\citealp{markarian1967, Smith1976, macalpine1977, wasilewski1983,
zamorano1994}} with the higher
sensitivity of a CCD detector. The survey method is described in detail by
\citet{salzer2000}. KISS selects objects for inclusion in the survey lists if
they possess a strong ($>$ 5$\sigma$) emission line in their low-dispersion
objective-prism spectra.  The survey has been carried out in two distinct
spectral regions: the blue portion (4800 -- 5500 \AA) where the primary line
observed is [\ion{O}{3}]$\lambda$5007, and the red region (6400 -- 7200 \AA)
where galaxies are selected by their H$\alpha$ emission.  
The current sample of galaxies were chosen from the \citet{jangren2005} list
of H$\alpha$ emission-line galaxy (ELG) candidates (hereafter KR3).
The objective prism passband imposes a redshift limit on the sample of $z \le
0.095$. The selection criteria used to define the sample were that the galaxies
exhibit spectra consistent with heating by star-formation processes (i.e.,
AGNs were excluded) and that they have a B-band absolute magnitude M$_B > -$18.0
(for H$_0$ = 75 km s$^{-1}$ Mpc$^{-1}$, although four are slightly brighter than
this limit when they are corrected for extinction). Thus these galaxies are
selected to be star-forming dwarf galaxies. 

These criteria produced a list of 26
galaxies within the NDWFS Bo\"otes area. However, of those 26 galaxies, only 19
overlap with the {\it Spitzer} Shallow Survey area because the {\it Spitzer}
survey field is smaller than the NDWFS field. We discuss only these 19 galaxies
in our analysis. Many of these galaxies also turn out to be low metallicity --
the median metallicity of the sample is $\log[O/H]+12 = 8.17$ (0.32Z$_{\odot}$).
With the exception of one super-solar and one
slightly sub-solar galaxy, they are all significantly sub-solar
($<0.6$Z$_{\odot}$, see \citealp{rosenberg2006} for sample details). 

All of the KISS ELG candidates in the Bo\"otes field possess higher dispersion 
follow-up slit spectra \citep{salzer2005b} that have been used to verify the 
reality of the putative emission lines seen in the objective-prism
spectra, calculate accurate redshifts, and to distinguish between the
various activity types that might be present in a line-selected sample (e.g.,
star-forming galaxies vs.~AGNs). These follow-up spectra provide us with a
great deal of useful information (e.g., accurate redshifts, emission-line fluxes
and line ratios, reddening and metallicity estimates). The combination of the
accurate B and V photometry from the original survey lists with these follow-up
spectra allow for the construction of a fairly complete picture of the
properties of the KISS ELGs.

Table 1 contains some of the optical parameters for the galaxies in this sample.
More details about the KISS data and parameters can be obtained from KR3.
\subsection{{\it Spitzer} Multiband Imaging Photometer Data}

The far-IR observations of the Bo\"{o}tes field were performed using
the ``medium scan'' mode of the MIPS instrument \citep{rieke2004}
on-board the {\it Spitzer Space Telescope}. This mode allows for efficient
coverage of large areas of sky with simultaneous observations at
24, 70, and 160$\mu$m. The MIPS detector at 24$\mu$m is characterized by a
resolution of 2$\darcs$45 per pixel on an array of 128$\times$128
elements; the 70$\mu$m detector has a resolution of 9$\darcs$98 per pixel on an
array of 32$\times$16 elements; and at 160$\mu$m, the effective
array operates with two rows of 20 16\arcs$\times$18\arcs\ pixels
separated by one row of non-functional pixels. The point spread
function (PSF) in the MIPS images is characterized by a full width at
half maximum of 6\arcs, 18\arcs\ and 40\arcs\ at 24$\mu$m, 70$\mu$m and
160$\mu$m respectively.  Data were reduced using the MIPS Data Analysis
Tool \citep{gordon2005}.  The astrometry of the final mosaic was
calibrated against the 2MASS survey \citep{jarrett2000} and is
accurate to 0.3\arcs\ rms.  The effective integration time per sky
pixel was $\sim$90\,s, $\sim$40\,s, and $\sim$8\,s at 24, 70,
and 160$\mu$m respectively.

At 24 $\mu$m 18 of the 19 galaxies were detected while 11 galaxies were detected
at 70 $\mu$m and only 3 were detected at 160 $\mu$m. Photometry was performed
with PSF fitting using the
IRAF/DAOPHOT software \citep{stetson1987}. At each wavelength an
empirical PSF was constructed from the brightest objects found in the
mosaic. The flux density of each source was derived from the scaled, fitted PSF 
plus a slight correction to account for the finite size of the
modeled point spread function. For the cases where residual
emission was found in the 24$\mu$m image after the PSF subtraction, aperture
photometry was performed within a region large enough to
account for the extended emission of the object. Only 5 sources show 
extended emission at 24$\mu$m and all of the targets appear unresolved at 70 and
160$\mu$m. For the non-detections at 70
and 160 $\mu$m, we adopted an upper limit based on the noise.
This noise was derived as the dispersion of the flux measured within
fixed-diameter apertures randomly placed over blank sky regions of the image.
The flux densities for all of the sources are presented in Table 2. 

\subsection{{\it Spitzer} 16 $\mu$m Imaging}

The galaxies in this sample were all imaged at 16$\mu$m with the blue peak-up
camera on the {\it Spitzer} Infrared Spectrograph (IRS\footnote{The IRS was a
collaborative venture between Cornell University and Ball Aerospace
Corporation funded by NASA through the Jet Propulsion Laboratory and
the Ames Research Center.}, \citealp{houck2004b, werner2004}). The targets were
observed as part of the IRS guaranteed time observing program on 2006
January 18 and 19. The 19 galaxies were observed with a 5 position random dither
pattern in order to oversample the point spread function (PSF) on the
$1.8\arcs\times1.8\arcs$ pixels of the IRS Short-Low Si:As detector. The
observing time was 30 seconds per position for a total on-source integration
time per target of 157 seconds. 

The data were processed by the
{\it Spitzer} Science Center pipeline (version 13.2). The 2D images were
converted from slopes after linearization correction, subtraction of
darks, and cosmic ray removal. The resulting images were divided by
the photometric flat and a world coordinate system was determined 
using the reconstructed pointing of the telescope. Final
rectified, shifted, co-added image mosaics were produced by the
pipeline. The astrometric accuracy of our images is better than
2$''$ and the FWHM of the PSF is 3.5$\farcs$ 

Photometry was
performed using fixed apertures of 3 pixels in radius. The flux
conversion factor was 0.01375 MJy\,sr$^{-1}$\,(e$^{-}$sec$^{-1}$)$^{-1}$,
and the aperture loss correction factor used was 1.38, as described in
version 2.0 of the IRS Data
Handbook\footnote{http://ssc.spitzer.caltech.edu/irs/dh/}. Our
photometric uncertainties are less than $\sim$6\% based on the uncertainty in
the photometric calibrators that were used. Ultimately 16 of the 19 
sources were detected in these images. The
flux densities for all of the galaxies are presented in Table 2.

\subsection{{\it Spitzer} Infrared Array Camera Data}

The majority of the NOAO Deep Wide Field in Bo\"otes was mapped in the 3.6, 4.5,
5.7, and 8.0 $\mu$m bands in January 2004 \citep{eisenhardt2004} using the
Infrared Array Camera (IRAC) aboard the {\it Spitzer} Space Telescope. 
The IRAC coverage of 8.5 square degrees was reached by tiling the $5\arcm \times
5\arcm$ field-of-view over the region. Each position in the survey field was
observed with
three 30 second IRAC frames, resulting in a depth of 19.1, 18.3,
15.9, and 15.2 Vega magnitudes ($5\sigma$) at 3.6, 4.5, 5.8, and 8.0$\mu$m,
respectively. A full discussion of the data and the reduction
procedures is presented by \citet{rosenberg2006}. All 19 galaxies were detected
in all four bands.

\subsection{Galaxy Evolution Explorer Data}

Archival ultraviolet (UV) imaging for all of the galaxies in this sample was 
available from the Galaxy Evolution Explorer (GALEX) \citep{martin2005}. 
These images were taken as
a part of the Deep Imaging Survey and span a wide range in exposure time. For
all of the systems, near-UV (NUV) images in the 1750--2800 \AA\ band are
available, and for a small subset (6) of the objects, far-UV (FUV) imaging in
the 1350--1750 \AA\ band is available. The data have been processed though the
standard GALEX reduction pipeline producing photometric images with an angular
resolution of 5$\darcs$6 (FWHM) in the NUV and 4'' (FWHM) in the FUV. 

Photometry was performed on the archival images using the ELLIPSE task in the 
STSDAS module to IRAF. The regions around the galaxies were first masked to
exclude the flux from neighboring sources. Most of the galaxies were then fit
with circular apertures (for the 3 distinctly elliptical galaxies, KISSR 2302,
2357, and 2359, elliptical apertures were used to minimize the noise). In order
to calculate the total UV counts for the galaxy, the size of the apertures was
allowed to grow until the background was reached as determined by the curve of
growth of the flux. The counts in each aperture were converted to fluxes using
the GALEX calibration
values\footnote{http://galexgi.gsfc.nasa.gov/FAQ/counts\_background.html}
\citep{morrissey2005}. The results are presented in Table 2.

\section{DUST DIAGNOSTICS}

\subsection{Mid-Infrared Emission from Star-Forming Dwarf Galaxies}

The existence of polycyclic aromatic hydrocarbon (PAH) emission from dwarf and
low metallicity galaxies has been discussed by several 
different groups (e.g., \citealp{wu2006, rosenberg2006, engelbracht2005,
madden2006, hunt2005, madden2000}). This subject is of increasing
interest because these molecules provide a probe of the star-formation
environment and, therefore, can be used as a measure of the star-formation rate
in galaxies.
\citet{rosenberg2006} showed that the [3.6]-[8.0] colors of these
galaxies span the full range of late-type galaxy colors. The presence of some
star-forming dwarfs with very red infrared colors indicates that even some metal
poor dwarf galaxies which have low line-of-sight internal reddening, as
measured by the c$_{H\beta}$ parameter, have a significant amount of dust.
Nevertheless, the only way to definitively determine whether a galaxy exhibits 
PAH emission is through infrared spectroscopy, but only a small number of
sources are bright enough to be measured in this way. 

To assess the degree to which PAH emission is present in the galaxies in this
sample, information from broadband flux ratios was combined with an examination 
of the galaxies' spectral energy distributions (SEDs; description of the SED
fitting procedures is presented in Appendix A). The fitting process provides a
comparison of the rough shape of the SED measured from the broadband data with a
set of galaxy templates. The relevant pieces of information are given in Table
\ref{tab:SED}. The general trends in the data are presented here while a
detailed description of the results for each galaxy in the sample are included
in Appendix B. The template fitting approach used here is not ideal
because it does not provide a physical measurement of the stars, dust, and PAHs
and because the templates are not derived from low metallicity galaxies. As has 
been shown by \citet{marshall2007} and \citet{galliano2007}, a proper analysis
of the dust, PAH emission and
stellar starburst component is not possible using only the IRAC and MIPS
photometric points even in ``simple" star forming galaxies  because of the
degeneracy in these component's contributions to the near-/mid-IR spectrum of
the galaxy. As a result, these templates provide only a general guide to the
properties of the galaxies and are used to answer two basic questions: (1) are
the SEDs of dwarf starbursts similar to those of ``normal" and starburst
galaxies? and (2) do dwarf starbursts have ``detectable" PAH emission? 

\begin{figure}
\epsscale{0.85}
\plotone{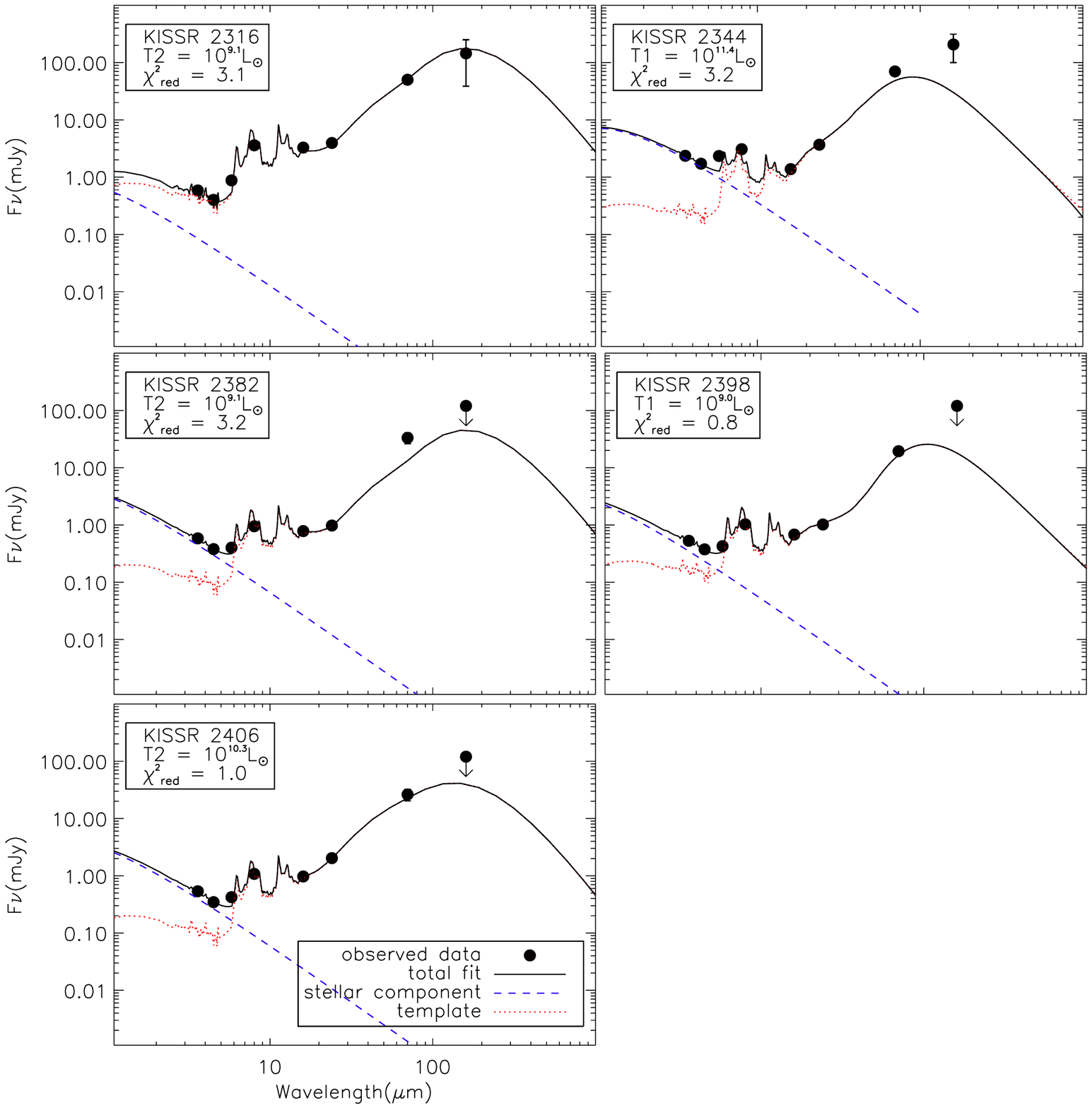}
\caption{SEDs for the sample galaxies that show evidence for PAH emission.  The
points represent the observations. The blue dashed line shows the best-fit 
blackbody representing the stellar component of the model. The dotted red line 
shows the best-fit dust template from the \citet{lagache2003} (T1) and
\citet{dale2002} (T2) models. The solid line is the sum of the stellar and dust
fits to the data. The values of T1 (Lagache templates) or T2 (Dale templates) in
the legend refers to the best-fit template -- it is a measurement of the shape
of the spectrum not of the infrared luminosity as it is not dependent on the
luminosity scaling. The dust templates are only fit longward of 8.0 $\mu$m
(i.e., not including the 5.8 and 8.0 $\mu$m points). The reduced $\chi ^2$
values for the fits are given in the legends.}
\label{fig:seds}
\end{figure}

\begin{figure}
\epsscale{0.85}
\plotone{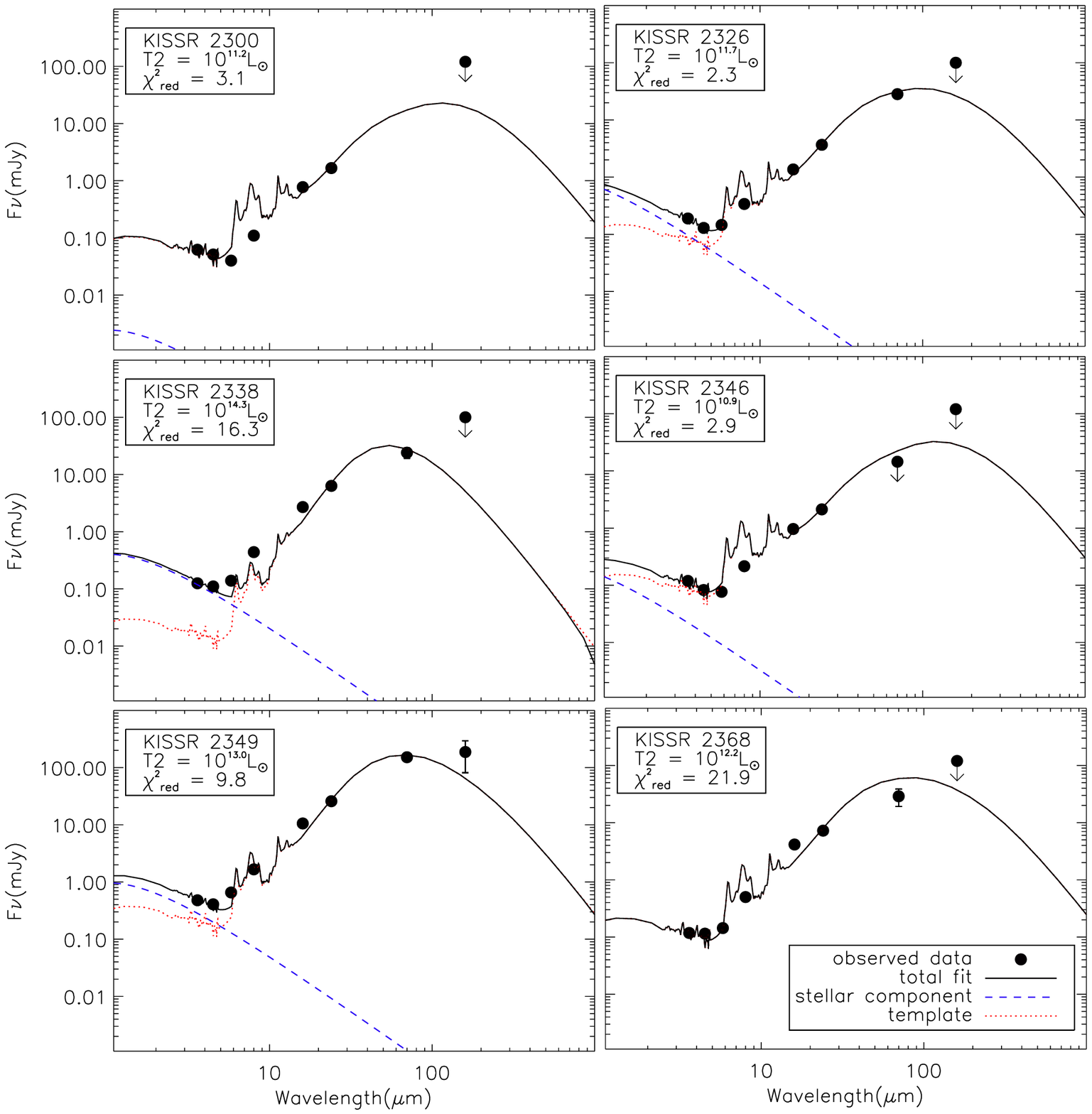}
\caption{SEDs for the sample galaxies that do not show evidence for PAH
emission.  The
points represent the observations. The blue dashed line shows the best-fit 
blackbody representing the stellar component of the model. The dotted red line 
shows the best-fit dust template using the \citet{dale2002} and 
\citet{lagache2003} models. The solid line is the sum of the stellar and dust
fits to the data. The values of T2 (Dale templates) in
the legend refers to the best-fit template -- it is a measurement of the shape
of the spectrum not of the infrared luminosity as it is not dependent on the
luminosity scaling. The dust templates are only fit longward of 8.0 $\mu$m
(i.e., not including the 5.8 and 8.0 $\mu$m points). The reduced $\chi ^2$
values for the fits are given in the legends.}
\label{fig:seds2}
\end{figure}

\begin{figure}
\epsscale{0.85}
\plotone{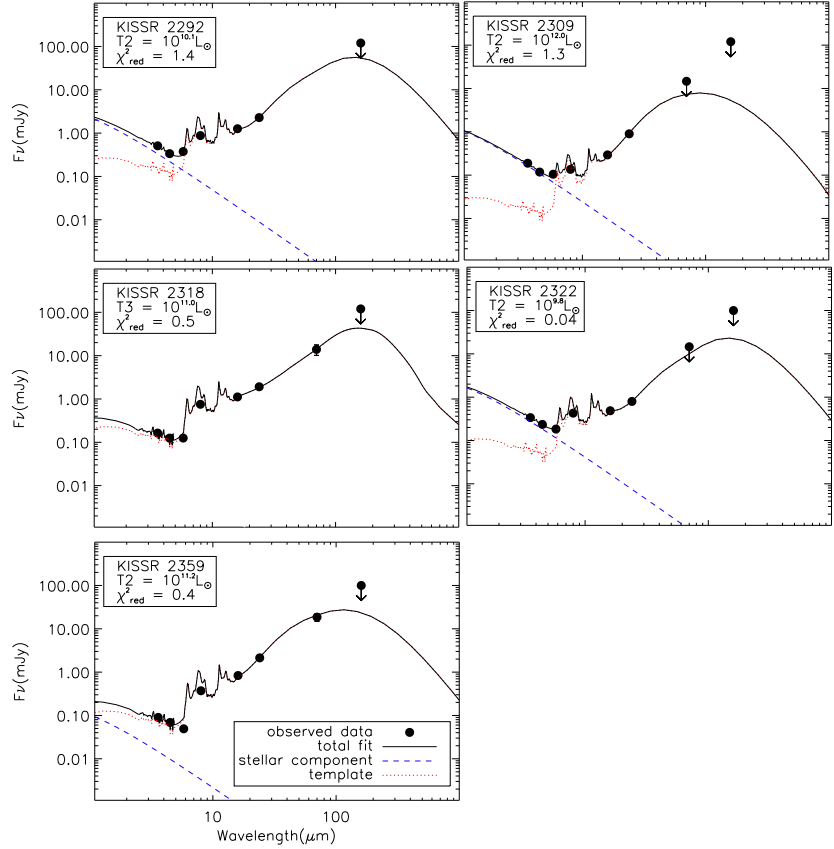}
\caption{SEDs for the sample galaxies for which a determination can not be made
as to whether PAH emission is present.  The
points represent the observations. The blue dashed line shows the best-fit 
blackbody representing the stellar component of the model. The dotted red line 
shows the best-fit dust template using the \citet{dale2002} and 
\citet{lagache2003} models. The solid line is the sum of the stellar and dust
fits to the data. The values of T2 (Dale templates) or T3 (normal galaxy
template) in
the legend refers to the best-fit template -- it is a measurement of the shape
of the spectrum not of the infrared luminosity as it is not dependent on the
luminosity scaling. The dust templates are only fit longward of 8.0 $\mu$m
(i.e., not including the 5.8 and 8.0 $\mu$m points). The reduced $\chi ^2$
values for the fits are given in the legends.}
\label{fig:seds3}
\end{figure}

\begin{figure}
\plotone{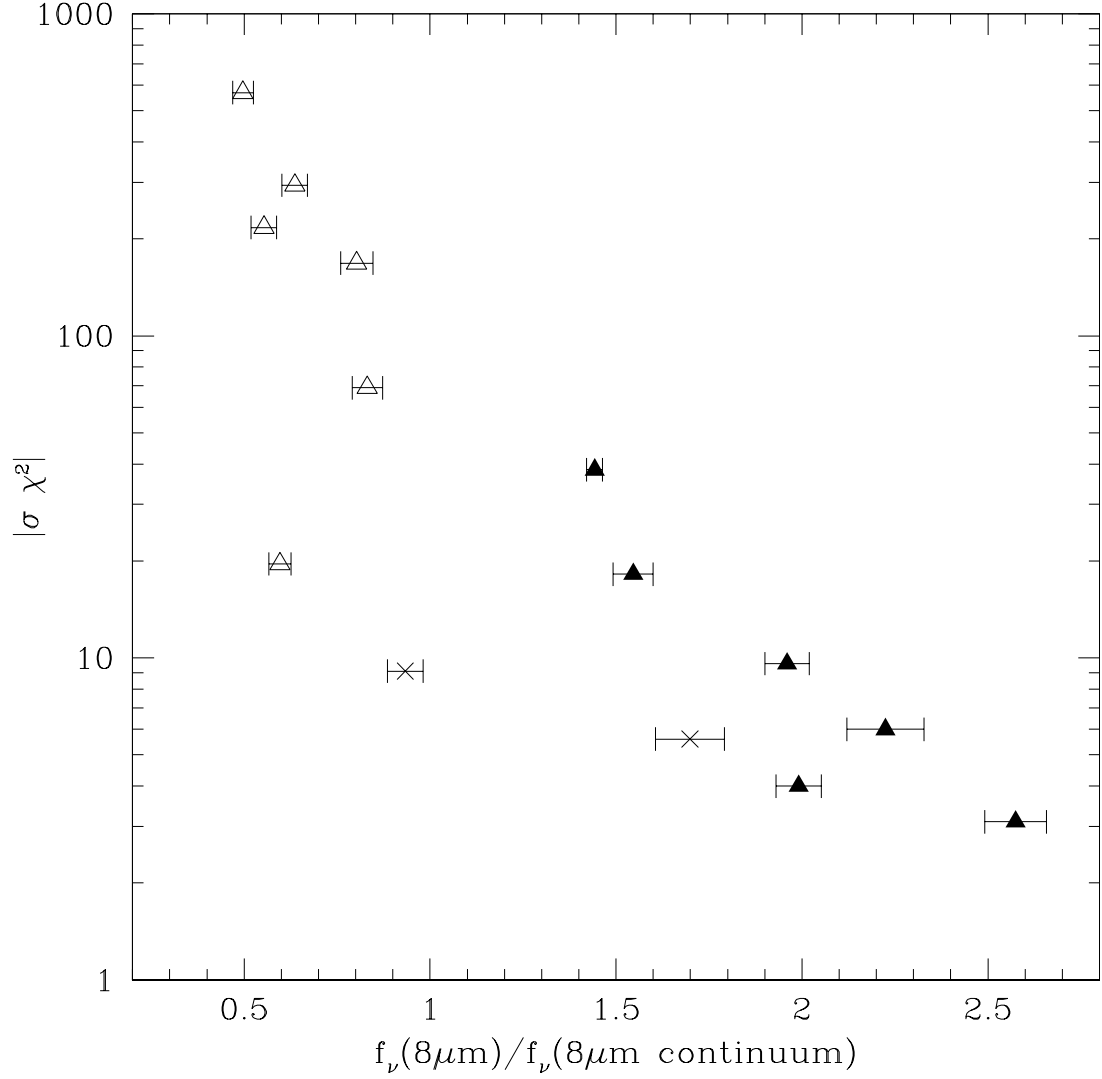}
\caption{The relationship between log$| \sigma \chi^2|$ and the ratio of the 8
$\mu$m flux to the 8 $\mu$m continuum. The y-axis represents a parameterization
of fit between the SED templates and the data -- it is the log of the absolute
value of the product of $\chi^2$
of the fit and the number of sigma difference between the fit and the
data at 8 $\mu$m. The 8 $\mu$m continnuum is an estimate based on a
straight line fit to the 5.8 $\mu$m and 16 $\mu$m data. Since the 5.8 $\mu$m
data can also have some PAH emission in it, this is a lower limit to the
continuum flux. Even though none of these dwarf galaxies have {\it Spitzer}/IRS
spectroscopy to determine the presence of PAH emission, we mark in filled/open
triangles the ones for which the SEDs indicate the presence/absence of PAH
emission. The galaxies for which the presence or absence of PAH emission could
not be determined are marked with $\times$s.}
\label{fig:8mratio}
\end{figure}

Figures \ref{fig:seds} to \ref{fig:vsg_diag} provide diagnostics of the PAH and 
dust emission properties of galaxies.
The model template fits to the galaxy spectral energy distributions (which were
not fit at 5.8 and 8.0~$\mu$m points where PAH emission has the largest impact)
are shown in Figures \ref{fig:seds}, \ref{fig:seds2}, and \ref{fig:seds3}. The
galaxies that have significant PAH emission (Figure \ref{fig:seds}) either
exceed the models at 8 $\mu$m or have 8$\mu$m fluxes that are within 3$\sigma$
of the models. Alternatively, all of the galaxies without PAH emission either
have large reduced $\chi^2$ for the template fits (the lowest $\chi_{red}^2 =
9.8$) or the templates exceed the data by at least 3$\sigma$ at 8$\mu$m. All of 
the other galaxies fall between these two criteria and are listed as having 
questionable PAH emission which probably means that they have a low level of PAH
emission. However, a definitive and/or quantitative
assessment of the PAH emission from this intermediate category can not be made
without spectroscopy. Note that there are also three galaxies for which SEDs are
not plotted because the long wavelength data necessary to constrain the fits do
not exist.

Figure ~\ref{fig:8mratio} shows the relationship between the quality of the fit
to the SED template and the 8$\mu$m to 8$\mu$m continuum ratio. The fit is
worse for large values of log$| \sigma \chi^2 |$. The ratio of 8 $\mu$m emission
to the 8$\mu$m continuum provides an alternate method for determining the
existence of PAH emission. Those galaxies that have detectable PAH emission have 
a flux ratio that is larger than one. This method of separating the galaxies is 
in agreement with the examination of the SEDs and would place KISSR 2359 in the
PAH detections category and KISSR 2309 in the non-detections category.

Figure ~\ref{fig:engplt1} shows the relationship between ratios of the 8~$\mu$m
emission and the dust continuum at shorter and longer wavelengths. The
division of galaxies into systems exhibiting PAH emission and those that do not
based on the relationship between the 8$\mu$m fluxes and the models as discussed
above agree well with the divisions established by \citet{engelbracht2005} based
on the relationship between the 8~$\mu$m emission and the dust continuum. The
8~$\mu$m excess above the dust continuum for
galaxies with PAH emission is reflected in the flux density ratio:
f$_{\nu}$(8.0)/f$_{\nu}$(24)$>0.3$.
The galaxies for which the SEDs do not definitively determine the nature of the
PAH emission populate both the ``PAH" and ``no-PAH" region of the plot.
KISSR 2292, 2318, and 2322 look like galaxies with PAH emission using this
measure while KISSR 2309 looks like galaxies without PAH emission and KISSR 2359
is on the boundary between the two regions.

Figure~\ref{fig:vsg_diag} shows how the ratio between the 8~$\mu$m emission and 
the hot dust continuum is related to the slope of the hot dust continuum. The 8
$\mu$m flux has been stellar subtracted assuming f$_{\nu}$(8.0)$_{stellar} =
0.232$f$_{\nu}$(3.6) (\citealp{helou2004}, the correction at 8$\mu$m is very
small and the stellar contribution at 16 and 24 $\mu$m is ignored because it is 
negligible.) The blackbody that was fit to the galaxy SEDs in \S 3.1 is not
used for this subtraction because the templates also include some stellar light
so this extrapolation from the 3.6 $\mu$m flux is more appropriate. In general
the galaxies that do not exhibit PAH emission have small
values for both of these ratios: f$_{\nu}$(8.0)/f$_{\nu}$(24)$<0.2$ and 
f$_{\nu}$(8.0)/f$_{\nu}$(16)$<0.3$ while all of the galaxies that exhibit PAH 
emission have f$_{\nu}$(8.0)/f$_{\nu}$(16)$>0.9$. The galaxies for which the SED 
fitting was not able to distinguish between PAH and no-PAH emission fall between 
these two groups in their f$_{\nu}$(8.0)/f$_{\nu}$(16) ratio and all fall below 
the starburst galaxy model curves. The PAH and no-PAH galaxies are better
separated in this plot than they are with the other MIR flux ratios. In
particular the galaxies are well separated using the
f$_{\nu}$(8.0)/f$_{\nu}$(16) ratio. 

For a small number of these galaxies, 70 and 160 $\mu$m data exist and the
models described in \citet{draine2007} can be used to  investigate a possible
explanation for the differences between the SEDs of these galaxies and those of
the templates. These models include PAH material
in addition to dust which consists of a mixture of carbonaceous and amorphous
silicate grains that have sizes consistent with the observed
wavelength-dependent extinction in the Milky Way while allowing the PAH
abundance and the radiation field to vary. While the smaller number of data
points inserted into the model should caution against over-interpretation of the
results, they do allow for a comparison of a few of our dwarfs with some more
``normal" galaxies. This comparison provides at least one explanation of the
differences we see between our galaxies and the model templates. The parameters
we derive from the model include q$_{PAH}$, the fractional contribution of PAHs
to the dust; a parameterization of the heating of the dust by starlight in the
intensity range between U$_{min}$ and U$_{max}$ given by: $dM_{dust} \over dU$
= (1-$\gamma$)M$_{dust} \delta$(U-U$_{min}$) + 
$\gamma$M$_{dust}$ ${(\alpha-1)} \over {(\Udiff)}$U$^{-\alpha}$; and f$_{PDR}$, 
the fraction of
the IR luminosity that is radiated in regions where U$>10^2$. We follow the
prescription in \citet{draine2007} for estimating these quantities for the
galaxies possessing enough data. For the galaxies without 160 $\mu$m detections
we use the 3 $\sigma$ value as an upper limit and obtain a lower limit by 
assuming that between 70 and 160 $\mu$m the galaxy's SED is Rayleigh-Jeans which
defines F$_{70}$/F$_{160} = 11.94$. 

For KISSR 2316 and 2344 which both have PAH
emission, U$_{min} \sim 2$, q$_{PAH} = $3.2\% and 1.8\%, $\gamma \sim$ 0.01 and
0.001, and f$_{PDR} = $ 7.6\% and 3.8\% respectively. For several other PAH
emitting galaxies with limits on the 160 $\mu$m flux the range of f$_{PDR}$ can
be computed. For KISSR 2406, f$_{PDR}$ = 5.8-6.9\% while for KISSR
2382 and 2398 the fraction is below the plot/fit range. Only a small fraction of
the dust emission in galaxies that exhibit clear PAH emission comes from high 
stellar intensity regions, according to these models.
The \citet{draine2007} dust diagnostics for KISSR 2349, which does not show PAH
emission, indicate that U$_{min} \sim 5$, q$_{PAH} = $1.12\%, $\gamma
\sim$ 0.09 and f$_{PDR} = $ 40\%. For galaxies for which we only have 160 $\mu$m
limits, KISSR 2326, 2338, and 2368, f$_{PDR} = $15-20\%, 31-35\%, and 30-34\%
respectively. The value of f$_{PDR}$ derived for these dwarf galaxies is much
higher than the values for the galaxies examined by \citet{draine2007}. This
difference in f$_{PDR}$ may indicate that these galaxies are experiencing the
destruction of PAHs in the photo-dissociation regions. These galaxies which do
not show PAH emission have model
parameters indicating a stronger radiation field and with more of the IR
luminosity originating in high intensity regions where it is more difficult for
the PAHs to survive. 

Why is the shape of high infrared luminosity templates of normal metallicity
galaxies more consistent with the observed SED of these low luminosity low
metallicity dwarf galaxies (specifically KISSR 2338, 2349, and 2368)?
One of the primary reasons that the high luminosity templates fit these objects
better is that they peak at shorter wavelengths and, in general, have a steeper
(redder) spectral slope in the 8 to 70 $\mu$m range which is also observed in
these dwarfs. In addition, the fits may be thrown off by minimal or absent PAH 
emission since all of the templates include the PAH features. This also explains 
why the fits to PAH-deficient sources are poor. 
A more physical reason for the poor fits may
be that for these compact galaxies the specific SFRs (normalized) by mass are
high so, like in the ultra-luminous infrared galaxies (ULIRGs), more of the PAHs
are being destroyed in the high intensity radiation. In particular it may be
that the area outside of the star-formation region, where most of the PAH
emission originates, is smaller in the galaxies that lack PAH emission because
the HII regions are beginning to overlap. 

\subsection{How do Dust Properties Relate to Metallicity?}

Metallicity is a parameter that can shape the properties of the interstellar
medium in galaxies and can span a wide range both within and between galaxies.
We explore how metallicity affects properties of the dust observable in the
mid-infrared emission from these systems. 

Figure~\ref{fig:abundrat} shows the variations in the ratio of the PAH
emission at 8.0 $\mu$m to the dust continuum at 24 $\mu$m with metallicity.
Both the galaxies from this sample and the blue compact dwarf galaxies 
\citep{wu2006} show a weaker correlation (correlation coefficient $\rho_{XY} =
0.36$) than the systems in the \citet{engelbracht2005} sample (correlation 
coefficient $\rho_{XY} = 0.75$). Star-forming dwarf galaxies span a wide
range in 8 to 24 $\mu$m flux density ratio -- over an order of 
magnitude at $\log[O/H]+12 = 8.2$. The transition from a
high flux density ratio to a low ratio at $\log[O/H]+12=8.2$, as claimed by
\citet{engelbracht2005}, appears to be more of a slow transition with a large
galaxy-to-galaxy variation at metallicities around this transition value.
Several of the galaxies in the $\log[O/H]+12=8.0$ to 8.2 range exhibit PAH
emission and there is nearly two orders of magnitude spread in the 8 to
24~$\mu$m flux density ratio in this metallicity range.

\begin{figure}
\plotone{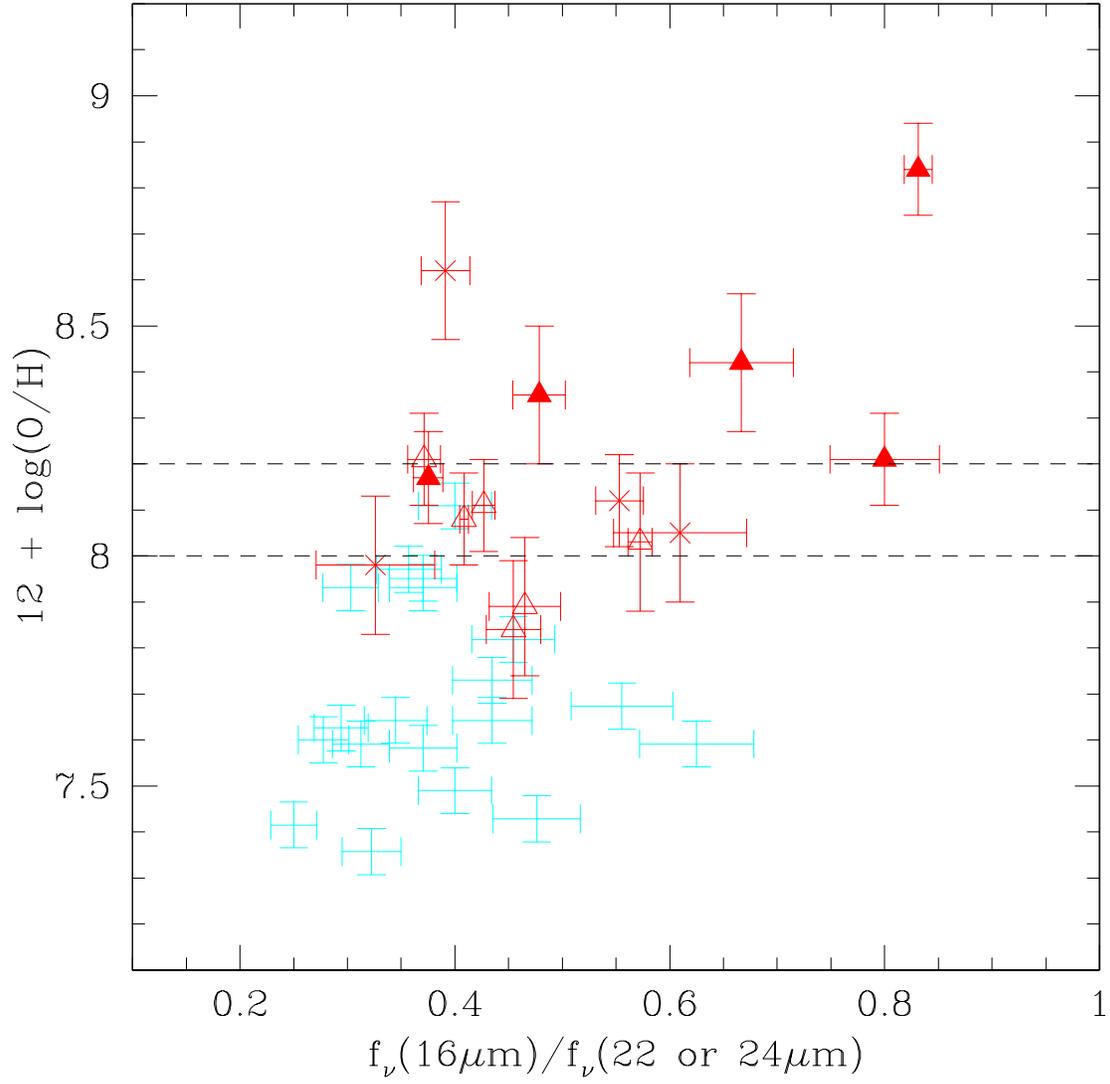}
\caption{Galaxy metallicity as a function of the 22 or 24 to 16 $\mu$m flux
ratio. The blue errorbars are blue compact dwarf galaxies from
\citet{wu2006} for which the presence or absence of PAH emission is unknown. The
red points are galaxies in this sample. Filled points are galaxies that show 
evidence for PAH emission while open points do not. The $\times$s are galaxies
for which the presence or absence of PAH emission could not be determined from
the SED fitting. The dashed lines 
delineate the region with metallicities between $\log[O/H]+12=8.0$ and 8.2}
\label{fig:abundrat2}
\end{figure}

Figure~\ref{fig:abundrat2} shows the relationship between 
metallicity and 16 to 24~$\mu$m flux density ratio for this survey
and the 16 to 22~$\mu$m flux density ratio for the blue compact dwarf
galaxies \citep{wu2006}. This flux density ratio is a measure of the continuum
slope in these galaxies. Fluxes measured at 24 and 22~$\mu$m should be
comparable because the bands are wide (5.5 and 7~$\mu$m respectively), the
spectrum is fairly smooth in this flux range, and the relative K-corrections
should be small -- for a standard dust SED, the 16 to 22~$\mu$m flux density
ratio would be slightly larger.  There is an increase in the flux density ratio
with increasing metallicity, but the spread at any given metallicity is large. 
For any given metallicity there is a wide range in the slope of the dust
continuum, but on average the lower metallicity galaxies have a steeper slope.
As mentioned previously, this steeper slope implies hotter dust which may be 
explained by higher radiation 
density and increased emission from very small grains in these dwarf galaxies.

Figures ~\ref{fig:abundrat} and \ref{fig:abundrat2} demonstrate that these 
galaxies exhibit a wide range of emission and/or dust properties at any given 
metallicity -- 
there is a wide range in the PAH to continuum ratio (Figure~\ref{fig:abundrat}) 
and in the continuum slope (Figure~\ref{fig:abundrat2}). 

\section{STAR-FORMATION RATE INDICATORS}

Star-formation rate plays an important role in the evolution of galaxies. As the
high redshift universe becomes more and more accessible, star-formation and its 
evolution can be studied over a wider range of epochs (e.g., \citealp{madau1996,
steidel1999,giavalisco2004}). Estimates of the evolution in the star-formation 
rate come from indicators that span the spectrum from the 
rest-frame UV (e.g., \citealp{madau1996,bouwens2006}) to the optical (e.g., 
\citealp{gallego2002, hogg1998}) to the infrared (e.g., \citealp{hwu2005,
calzetti2005}). All of the SFR indicators commonly used make assumptions about
the metallicity, stellar make-up, star-formation mode (continuous versus
instantaneous star formation) and/or interstellar medium properties of the 
systems. In general the conversion from flux to star-formation rate assumes a 
standard IMF in a solar metallicity galaxy with interstellar medium properties 
consistent with a ``normal" galaxy in the local universe. In order to use these
indicators at higher redshift, where low metallicity star-forming 
galaxies may be more prevalent, the connection between flux and SFR must be 
evaluated for a range of metallicity and galaxy type. Because it contains a
large fraction of 
low metallicity systems, this sample can be used to compare the affect of galaxy
metallicity on the SFRs calculated at different wavelengths.

\subsection{Description of Star-Formation Rate Indicators}

{\it Spitzer} is now providing an unprecedented view of galaxies in the MIR over
a wide range of redshifts. At these wavelengths, PAH emission and the dust
continuum are the primary contributors to the flux. The dust emission at 12
$\mu$m was first shown to be linearly correlated with a galaxy's bolometric
luminosity by \citet{spinoglio1995} using data from IRAS. More recently
correlations have been found between 15 $\mu$m emission and the total infrared
luminosity (TIR, \citealp{chary2001}), between both 8 and 24 $\mu$m emission and
the radio continuum emission from galaxies \citep{hwu2005}, and between
combinations of 8 and 24 $\mu$m emission and TIR \citep{calzetti2005}. 
In the ultraviolet (UV), GALEX \citep{martin2005} is opening up another window 
on star formation in the local universe and providing a probe of the emission 
from the youngest and hottest stars in galaxies. This section is devoted to a
comparison of star-formation rates derived using indicators from these different
wavelengths ranges. Several of the methods are obviously inappropriate for these
galaxies, but the goal is to understand the errors in the calculated SFR using 
standard techniques so we apply them for the sake of comparison.

\begin{itemize}

\item Several authors have built libraries of galaxy templates that can be used
to estimate TIR from the 24 $\mu$m flux densities 
\citep{dale2001,chary2001,dale2002,lagache2003,chanial2003} alone. For this
measurement only the 24 $\mu$m flux density is used to predict the total
infrared emission, not the full SED shape. As shown by
\citet{lefloch2005}, these templates all produce similar results; we adopt the
\citet{chary2001} results here without concern that this will bias our findings
significantly.  A single template was derived for each
luminosity bin (see \citealp{chary2001} for details) and can be used to predict
the TIR from the 24 $\mu$m luminosity. The \citet{kennicutt1998b} conversion was 
then used to derive the SFR from TIR. The galaxies
that were used to derive these templates are Arp 220, NGC 6090, M82, and M51
which were selected to represent ULIRGs, LIRGs, ``starbursts," and ``normal
galaxies" respectively. An additional set of far-infrared templates from
\citet{dale2001} were added to span a wider range of spectral shapes. 
In \S 3.1 we used a similar set of templates from \citet{lagache2003} and 
\citet{dale2001} to fit
the sources in this sample and found that, in many of the cases, the shape of
the low luminosity templates was not appropriate for fitting the galaxies. The
difference between the lwo luminosity template shape and the galaxy SED will
introduce an error in determining the SFR. 

\item \citet{hwu2005} used a sample of star-forming galaxies in the {\it
Spitzer} First Look Survey to study the correlations between 8 $\mu$m and 24
$\mu$m luminosity with 1.4 GHz and \ha\ luminosities. These correlations were
used to derive the 1.4 GHz luminosity from either 8 or 24 $\mu$m luminosities
which was then used to calculate SFR using the \citet{kennicutt1998b}
conversion. Most of the \citet{hwu2005} systems are
normal galaxies but a few dwarfs are included and appear to show a different
slope between the MIR and radio luminosity. The difference in the normal and
dwarf galaxy slopes
is probably driven by a lack of PAH emission (Figure \ref{fig:seds} and 
\ref{fig:seds2}) and the wide range of continuum slopes (Figure
\ref{fig:abundrat2}). The more recent calibration of the conversion between 24
$\mu$m luminosity and SFR from \citet{calzetti2007} does not produce significantly 
different results than the \citet{calzetti2005} results plotted here.

\item \citet{calzetti2005} showed that there is a correlation between 
$\log[\nu L_{\nu}(24)/{\rm TIR}]$ and $\log[L_{\nu}(8)/L_{\nu}(24)]$ for
\ion{H}{2} regions in NGC 5194. The SFR was calculated using the 
\citet{kennicutt1998b} conversion from TIR. \citet{calzetti2005} use 
$\log[L_{\nu}(8)/L_{\nu}(24)]$ instead of just 8 or 24 $\mu$m luminosity because
the scatter from the ratio is lower. However, there are still region-to-region
differences in the correlation between flux density ratio and SFR for the 
\ion{H}{2} regions in NGC 5194, all of which are high metallicity (they are 
consistent with $2\la Z \la3$~$Z_{\odot}$) so
the scatter is probably worse when studying objects that cover a range in 
metallicity. More recent

\item The total infrared star-formation rate has been calculated using 
the 24, 70, and 160 $\mu$m data for the three galaxies for which
all of these data are available using the method for deriving TIR described by 
\citet{dale2005}. The SFR was then calculated using the \citet{kennicutt1998b} 
conversion from TIR. For the galaxies for which 24 and 70 $\mu$m data are 
available but the 160 $\mu$m observation was a non-detection (an additional 8 
galaxies), we use the 3 $\sigma$ upper limit at 160 $\mu$m to place an upper 
limit on the total infrared luminosity. We also obtain a lower limit on the 
luminosity of these galaxies by assuming that between 70 and 160 $\mu$m the 
galaxy's SED is Rayleigh-Jeans which defines F$_{70}$/F$_{160} = 11.94$. 

\item Star-formation rates for all of the galaxies in this sample have been 
computed from the \ha\ line fluxes measured from the KISS objective prism 
spectra. These spectra reflect the total \ha\ emission from 
the galaxy because they are derived from objective prism observations.
A full description of the line flux measurements and calibration is given by
\citet{jangren2005}. In order to compute the star-formation rate for these
galaxies from the \ha\ flux, the measurements must be corrected for the
presence of blended [\ion{N}{2}] emission
(these are low resolution spectra), and for absorption. These corrections are
derived from slit spectra. The absorption
correction is derived using the reddening coefficient derived from the Balmer
line ratios, c$_{H\beta}$. From the [\ion{N}{2}] and absorption-corrected \ha\
fluxes, the SFRs were computed using the standard prescription from
\citet{kennicutt1998}.

\item In addition to using the standard methods for computing SFR from the \ha\
emission, we use a method that takes into account the harder radiation field in
low metallicity galaxies \citep{lee2002}. The stellar population synthesis
models on which the standard conversion assume a softer radiation field 
than what exists in these low metallicity systems. This prescription was derived
for a sample of KISS galaxies for which the selection criteria were the same as
for these objects. The resulting star-formation rates are lower than what is
obtained using the standard models because the harder radiation field produces
more \ha\ emission for a given star-formation rate. 

\item The UV emission from galaxies comes from the youngest, hottest stars. 
\citet{kennicutt1998} provides a conversion from the UV luminosity to SFR which
can be computed across the UV-band because young stars produce a flat spectrum
in this region. The UV SFR for these galaxies has not been corrected for
absorption so it is a lower limit which measures only the amount of
star formation not buried behind dust. Therefore, the SFR measured this way 
is complementary to that measured in the MIR where all of the emission is 
coming from the re-radiation from dust in the galaxy. We compute the total SFR
as described in \citet{bell2005} as: $\Psi /$ M\solar\ yr$^{-1} = 9.8 \times 
10^{-11}$ (L$_{IR} + 2.2$L$_{UV}$). We use the \citealp{calzetti2005} measurement 
of L$_{IR}$ because the values are generally between the other MIR results. As
with the infrared measurement, the sum of the MIR and UV measurement could 
overestimate the SFR if some of the PAH emission is excited by an older stellar 
population.

\end{itemize}

The SFRs computed using all of the above methods are reported in Table 3.

\subsection{Comparison of Star-Formation Rates}

\begin{figure}
\plotone{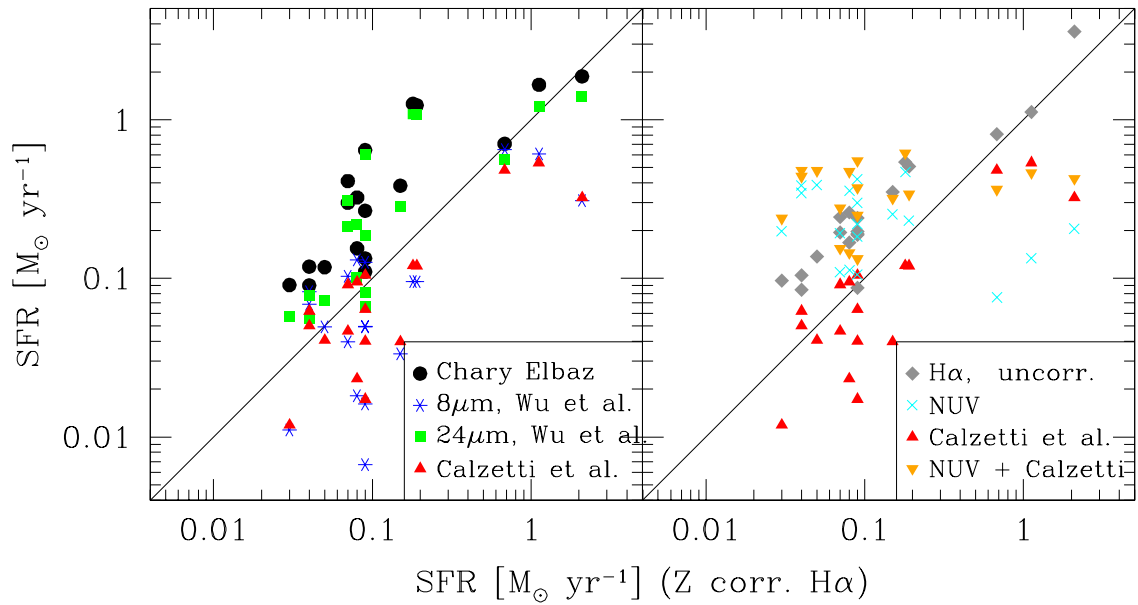}
\caption{Comparison of SFR indicators with respect to the SFR
calculated from \ha\ with a correction for metallicity. The left-hand panel
shows a comparison of the MIR indicators while the right-hand panel shows the
\ha\ not corrected for metallicity, the \citet{calzetti2005} MIR values, the NUV
measurement, and the combined NUV and MIR values. Because of a difference in the
calibration between the \citet{calzetti2005} MIR SFR and the \citet{bell2005}
combined MIR and NUV SFR, there are a couple of galaxies for which the 
\citet{calzetti2005} MIR SFR is slightly higher than the combined SFR.}
\label{fig:MIRSFR}
\end{figure}

Do all measurements of SFR give the same answer and, under what conditions are
the derived SFRs comparable? The answer to this question is critical if
measurements of SFR from the MIR and UV are going to be used along with \ha\ to
compute SFR density over a wide range of redshift. Figure \ref{fig:MIRSFR} shows 
the relationship between the MIR measurements of SFR and the
metallicity-corrected \ha\ SFR (left-hand panel) and between SFR measured over a
wide range of wavelengths and the metallicity-corrected \ha\ SFR (right-hand
panel). The errors due to the galaxy photometry on all of these these data
points (and those in the subsequent SFR plots), with the exception of the NUV
values, are smaller than the points. The NUV SFRs are the exception because the
values are all lower limits since no correction
has been made for extinction. There is also a distance error associated with
these measurements but they will only shift all of the points up or down without
altering their relative separations (i.e., it will not change
the spread between the points). These plots show that there is over an order of
magnitude spread in the SFR measurement for a given galaxy and the disagreement
occurs for both high and low SFR sources. However, the plot also
shows that several of the methods give similar answers. The \citet{chary2001}
and the 24 $\mu$m \citet{hwu2005} values are similar as are the 8 $\mu$m 
\citet{hwu2005} and the \citep{calzetti2005} values. These values agree because
they were derived in a consistent way for normal galaxies and are derived from
the same fluxes (24 $\mu$m and 8 $\mu$m respectively). The spread in the derived
SFRs points to a systematic difference in the calculation of these values due
either to the intrinsic spread in the luminosity-SFR relationships or to
a physical difference between these galaxies and those for which the
luminosity-SFR relationship was derived.

In general, the 
measurements based on the 24 $\mu$m luminosity (\citealp{chary2001} and 24 
$\mu$m \citealp{hwu2005}) tend to indicate a higher SFR than the 8 $\mu$m 
\citep{hwu2005} or 8 plus 24 $\mu$m \citep{calzetti2005} measurements. This 
result is consistent with the 8 $\mu$m to \ha\ correlation having a steeper
slope for dwarf galaxies than it does for more luminous galaxies as was found in
the \citet{hwu2005} study. 
For six galaxies in this sample the SFR derived from the 8 $\mu$m flux is about
an order of magnitude lower than that derived from the 24 $\mu$m flux. All six of
these galaxies are low metallicity systems (Z \lapprox 0.25 Z$_{\odot}$) that 
have SEDs which show no evidence for PAH emission. KISSR 2359 which has low
PAH emission, has an 8
$\mu$m derived SFR that is 4.5 times smaller than the 24 $\mu$m SFR. Clearly the
8 $\mu$m emission is not a good SFR indicator for low metallicity objects that
may not exhibit
PAH emission. Alternatively, the 24 $\mu$m SFR seems, in general, to overpredict 
the SFR in these galaxies relative to the \ha\ measurement. The galaxies in this
sample are, in general, very compact
systems with hard radiation fields because they tend to also be low
metallicity systems. These conditions appear to be driving the heating of hot
dust and/or very small grains that dominate in the region between 8 and 24 $\mu$m 
as has been seen previously in low metallicity systems \citep{madden2006}. The
difference between the 24~$\mu$m and \ha\ SFRs is certainly affected by the
incorrect calibration of the 24~$\mu$m SFR for these galaxies which comes about
because the SEDs shown in \S 3.1 are not well fit by the low luminosity
templates. 

The right-hand panel in Figure \ref{fig:MIRSFR} shows the SFR computed using
several different indicators including the \citet{calzetti2005} MIR indicator
and the NUV emission. The
SFRs computed in the NUV have not been corrected for extinction. For the highest
SFR systems (which are also the highest metallicity ones), the lack of an
extinction correction has led to a low measurement of the SFR. However, for
most of the systems, the SFR computed in the UV is larger than that computed
from the \ha\ despite the fact that extinction has not been taken into account.
One factor in the high UV SFR may be that the there is a metallicity dependence
that has not been accounted for. However, the significant excess in the UV SFR
even for galaxies that have metallicities $\gapprox 0.5$Z\solar, indicates that
this is probably not the only factor. 
A higher SFR measured in the UV than in \ha, attributed to declining
star-formation from a young but quasi-stellar burst, has also been seen for
intergalactic star-forming regions in NGC 5291 \citep{boquien2007}. A declining
SFR in the KISS galaxies would explain the discrepancy between the \ha\ and UV
SFRs since the timescale in the UV is about 10 times longer than it is in \ha.
However, this timescale argument would not explain why the UV SFR is higher than
the MIR SFR which does not decline as quickly. A different explanation would be
that the interstellar medium has been blown out in these systems so that there
is much less gas and dust surrounding the star-forming region. This explanation
is consistent with the mid-infrared spectroscopic results for a similar sample
of galaxies which provides evidence for supernova driven bubbles surrounding the
star-formation regions \citet{ohalloran2007}. A much more 
detailed analysis would be required to assess whether there was a strong burst
of star-formation in the recent past that could have blown away the gas 
and dust. The plot includes the combined MIR and NUV SFR using the
\citet{bell2005} formulation. For two of the galaxies, the total SFR is smaller
than the \citet{calzetti2005} MIR SFR alone because there is a small difference
in the calibration in the two methods. This calibration uncertainty is a
systematic error in all of the calculations and these results are consistent
within that error.

\begin{figure}
\plotone{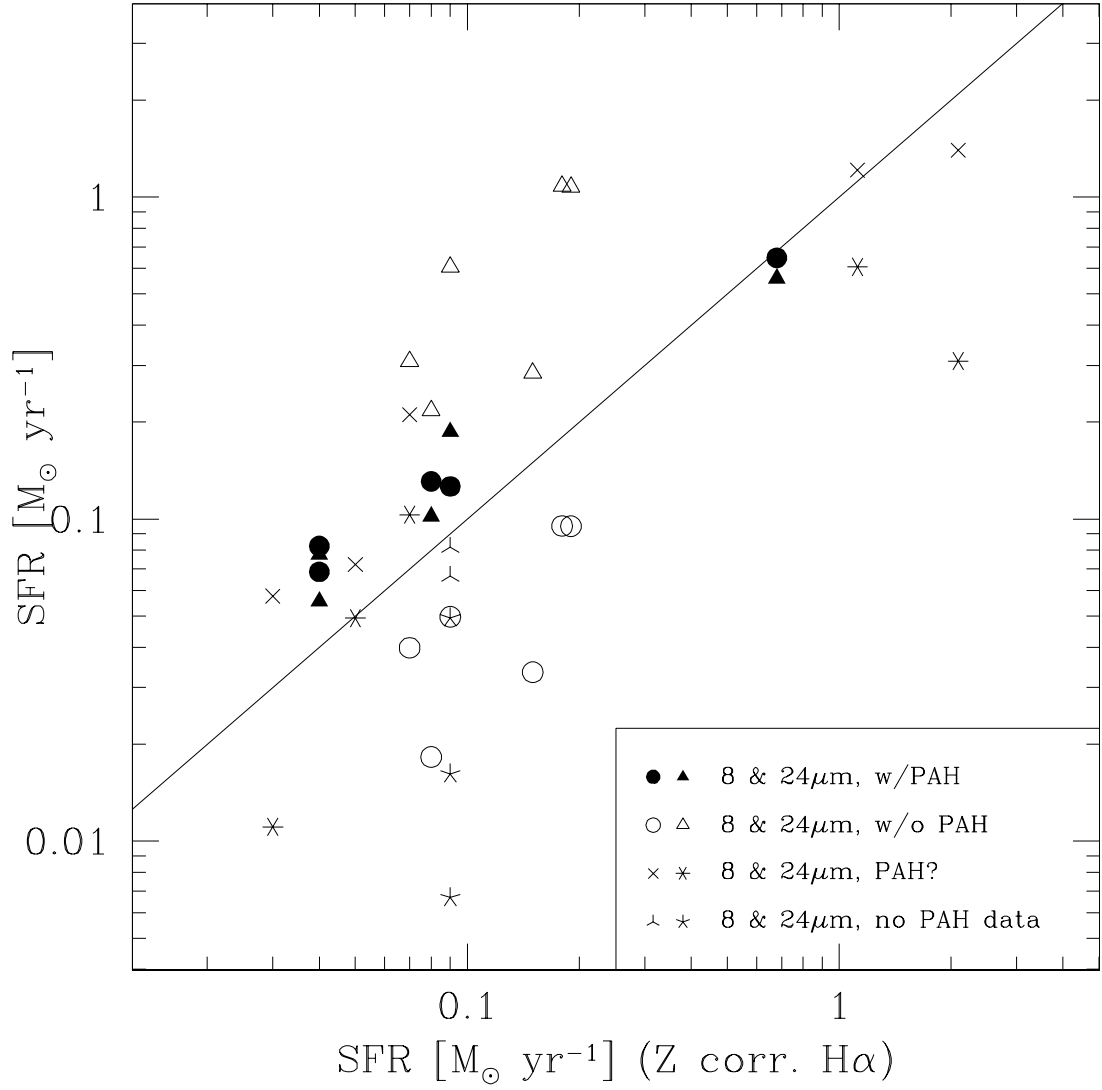}
\caption{Comparison of MIR SFR indicators as a function of the metallicity
corrected \ha\ SFR. The open symbols indicate objects that do not show evidence
for PAH emission while the filled symbols indicate those which do have PAH
emission. The crosses and stars indicate galaxies for which the PAH
determination was uncertain from the SED fitting. The ``no PAH data" points are
galaxies for which the longer wavelength data were not available so no SED
fitting could be done.}
\label{fig:SFRplot}
\end{figure}

Figure \ref{fig:SFRplot} shows the MIR SFR as a function of the \ha\ SFR with
point type specifying those galaxies with and without PAH emission.  For the
galaxies that appear to have PAH emission, the 8 and 24~$\mu$m measurements
of SFR are, in all but one case, in good agreement with the \ha\ values. For  
galaxies that do not show evidence for PAH emission, the SFR computed from the 8
$\mu$m flux is lower than what is measured from the \ha\ emission, while the
value computed from the 24 $\mu$m emission is higher in all but one case.  The
dispersion in the values of SFR computed in the mid-infrared for these galaxies
appears to be driven by a difference in the properties of the interstellar
medium in these galaxies. Both the PAH emission and the shape of
the MIR continuum differ from those of normal galaxies both in a lack of PAH
emission and in excess emission from hot dust and/or small grains. 

\begin{figure}
\plotone{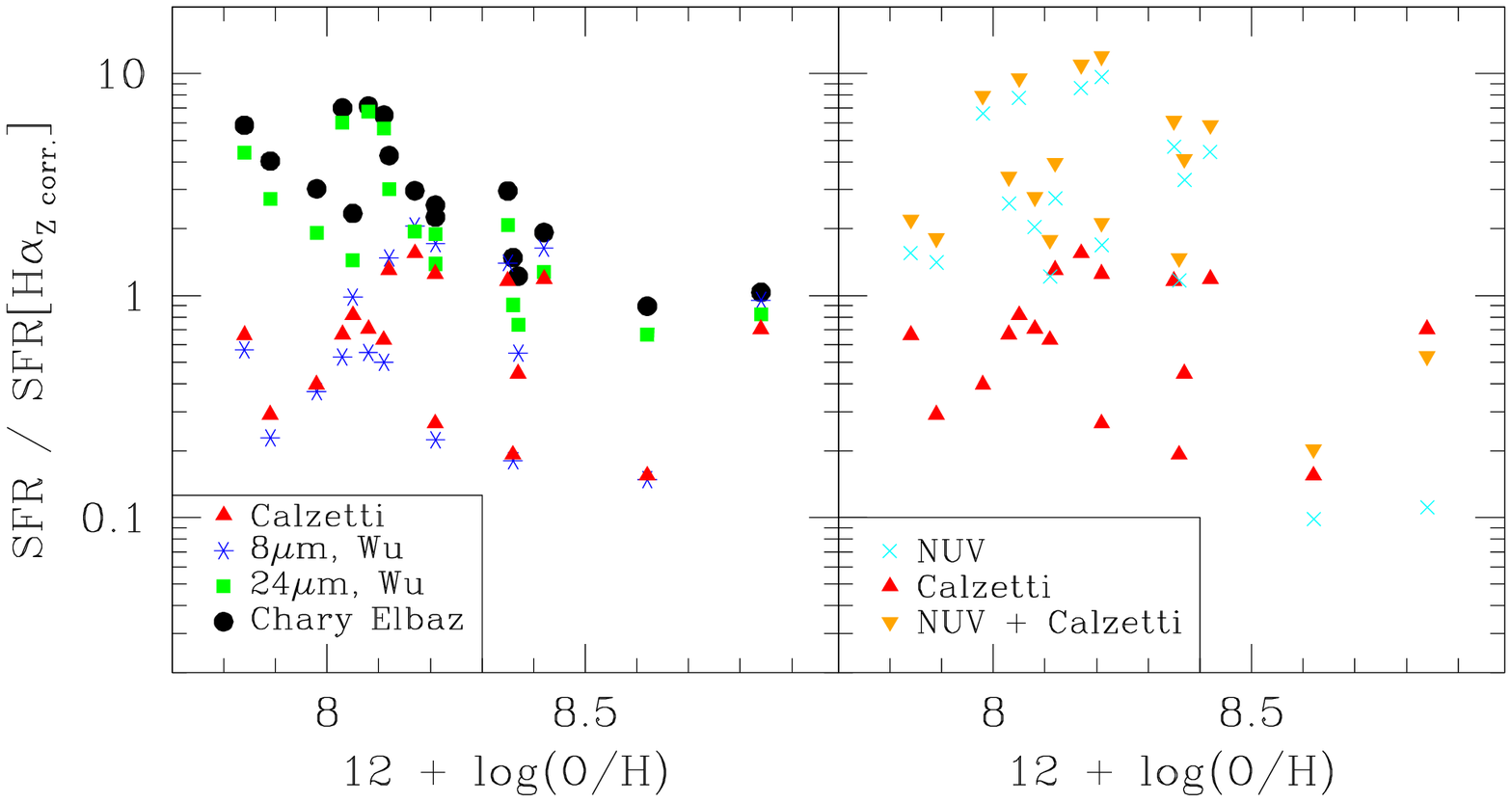}
\caption{Comparison of SFR indicators, normalized by the \ha\ corrected SFR as a
function of metallicity. The
left-hand panel is a comparison of the MIR indicators. The right-hand panel 
is a comparison of the \citet{calzetti2005}, the NUV, and the combined NUV and
MIR values. Because of a difference in the
calibration between the \citet{calzetti2005} MIR SFR and the \citet{bell2005}
combined MIR and NUV SFR, there are a couple of galaxies for which the 
\citet{calzetti2005} MIR SFR is slightly higher than the combined SFR.}
\label{fig:SFRmet}
\end{figure}

Figure \ref{fig:SFRmet} shows the relationship between the SFR ratio -- SFR from
a given indicator divided by the metallicity corrected \ha\ SFR -- as a function
of metallicity. The left-hand panel shows this ratio for the MIR indicators. The
right-hand panel shows the ratio for the \citet{calzetti2005} MIR SFR, the NUV
SFR, and the combination of the MIR and NUV SFRs. For the MIR SFRs, the two
measures that are derived from the 24 $\mu$m emission are the most highly
correlated with metallicity. For the 8 $\mu$m computed SFRs, a trend with
metallicity is not clear in these data. The NUV SFR for the highest metallicity 
systems are low because of extinction, but there is no clear trend among the
lower metallicity systems. 

This sample has 3 galaxies that have the properties of dusty starbursts
-- KISSR 2316, 2318, and 2359. These systems all have SFRs calculated from the
total infrared emission  that are much higher than that calculated from the NUV
(which is not corrected for extinction). These galaxies are some of the most 
luminous systems in the sample and the two with known metallicities are the 
highest metallicity systems. Since the derivation of the SFR indicators in
the mid-infrared has generally been based on dusty starbursts one might expect
the most consistency in the derived SFRs from these systems. The rates
calculated from \ha, 8 $\mu$m, 24 $\mu$m, and the total infrared are fairly
consistent for KISSR 2316 and 2318 (the UV values are low because they are not
corrected for extinction). However, the SFRs computed for KISSR 2359 are more 
problematic. This source has the highest \ha\ SFR, (3.59 \msunyr; 
not correcting for metallicity because the source is close to \zsun ) in the 
sample. The 24 $\mu$m SFRs from the \citet{chary2001} models and from the 
\citet{hwu2005} correlation are 1.87 and 1.40 respectively. These are smaller
than the \ha\ values, but given the spread in the
models and correlation they are still consistent. However, the 8 $\mu$m
SFR from the \citet{hwu2005} correlation is only 0.31, significantly less than
what is measured from the other methods. The lower value for the 8 $\mu$m
emission may be a reflection of a substantial portion of the infrared emission
coming from a high intensity region where the PAHs are destroyed.

\section{SUMMARY}

Star-forming dwarf galaxies have a wide range of dust properties that are quite
different from those of the spiral and massive starburst galaxies. These 
differences most likely result from a combination of the harder radiation field, 
the more compact structure of these systems, possibly a lack of dust due to
destruction in shocks or delayed injection, and either a different composition, 
different heating, or different distribution of the dust in the interstellar
medium.

We have used broadband SEDs and color-color diagrams to determine the presence or 
absence of PAH emission in the sample galaxies. It appears that one of the best 
indicators of PAH emission may be the 8 to 16$\mu$m flux ratio. Mid-infrared
spectroscopy, which is not available at this time, is required to confirm these
results. Nevertheless, the combination of the SEDs and
the color-color plots provide more information on the presence or absence of PAH 
emission than the trend of color with metallicity that was evident from only the 
{\it Spitzer}/IRAC data \citep{rosenberg2006}. 

One of the most striking properties of the
galaxies in this sample is that, despite their simplicity in being compact
systems dominated by a small number of \ion{H}{2} regions, they have a wide
range of infrared properties. There are low metallicity systems that show some
evidence for PAH emission (KISSR 2322, Z=0.24Z\solar) and high metallicity
systems for which PAH emission is uncertain (KISSR 2359, Z=0.91Z\solar)
indicating that metallicity is not the only parameter driving the presence
and/or strength of the PAH emission features. 

It is not only the PAH emission features in these star-forming dwarf galaxies
that are different from the spirals and massive starbursts, but also the shapes
of the mid-infrared continuua. Many of these galaxies have steep, red
mid-infrared continuua similar to those observed by \citet{gallais2004} in
starburst galaxies, by \citet{madden2006, hunt2006} in low metallicity galaxies
and modeled by \citet{galliano2005} as being due to emission from very small
grains.

Applying the model of \citet{draine2007} provides one possible interpretation of
the differences between the SEDs of the dwarf galaxies and the model templates.
The galaxies for which the mid-IR flux ratios indicate a lack of PAH emission
have a large fraction of their infrared emission coming from the
photodissociation regions (15-40\%) relative to galaxies for which the mid-IR
flux ratios indicate the presence of PAH emission (2-8\%). PAHs are more likely
to be destroyed in the high intensity photodissociation regions which would
explain why the PAH fraction is lower where these regions dominate. A physical
model that might explain this circumstance would be that the \ion{H}{2} regions
are beginning to overlap thereby shrinking the relative volume of
the lower intensity regions.

The effect of metallicity on the star-formation rate in galaxies is well studied
\citep{lee2004,kewley2004,moustakas2006}, primarily with respect to the optical
indicators. At 24 $\mu$m we see a clear connection between the derived
star-formation rate and metallicity, but the connection is not as clear at 8
$\mu$m where other factors must be dominating the dispersion in the relationship
between flux density and star-formation rate.

The differences between these star-forming dwarf galaxies and the majority of
systems that have been studied in the infrared have two major consequences: (1) 
mid-infrared spectral templates do not provide a good fit to the systems, and
(2) the methods that have been developed to determine the star-formation rate
from mid-infrared observations do not work well in these systems. Because these
galaxies are different from the majority of galaxies that have been studied in
the mid-infrared, some caution needs to be exercised in fitting spectral
templates and deriving mid-infrared star-formation rates, particularly at higher
redshift where more information is often not available and where compact
star-forming systems may be more prevalent.

\acknowledgements
This work is based in part on observations made with the Spitzer
Space Telescope, which is operated by the Jet Propulsion Laboratory,
California Institute of Technology under a contract with
NASA. Support for this work was provided by NASA through an award
issued by JPL/Caltech. JLR acknowledges support from NSF grant AST-0302049.
ELF's work was supported through NASA/JPL by the Spitzer Fellowship program.
JJS gratefully acknowledges support for the KISS project from the NSF through
grants AST 95-53020, AST 00-71114, and AST 03-07766.
VC would like to acknowledge partial support from the EU ToK grant 39965.

\appendix

\section {Description of SED Fitting Procedure}

The spectral energy distributions were studied by fitting model templates to
each galaxy. The 5.8 and 8.0~$\mu$m points which are most affected by PAH
emission were excluded from the fitting so that we could compare the photometry 
with the models at these wavelengths. Note that $\sim$10\% or less of the
8.0~$\mu$m emission in these galaxies is from stellar photospheres, most of the 
emission is from the dust continuum or PAH features.

The models consist of an ensemble of starburst galaxy model templates produced
by \citet{lagache2003} and normal star-forming galaxy templates by
\citet{dale2002}. Each template is identified by a galaxy luminosity. However,
the templates are used to model the shape of the SED and are not necessarily 
dependent on the luminosity of the source (we ignore the luminosity scaling).
These templates sample the MIR colors of both normal
and ultra-luminous infrared galaxies. The \citet{lagache2003} models also
include a single ``normal" (non-starburst) galaxy template. For many of the
galaxies, the fits are constrained only by the 16 and 24
$\mu$m data points.

In order to fit the broadband data, ``synthetic" data points for the
IRAC 3.6, 4.5, IRS 16$\mu$m, and MIPS 24, 70 and 160$\mu$m bands were created by 
convolving the template spectra with the filter response functions. To fit the 
near-infrared portion of the SED, a blackbody spectrum with a temperature in the 
range 5000 K to 10000 K was added to the template model. A least squares
technique was then used to determine the best fit template plus blackbody (steps
of 1000 K were used to determine the best-fit temperature) with the appropriate
scaling factors. The temperature of the chosen blackbody did not have a
significant effect on the SED fit. 

Out of the 19 sample galaxies, three are not detected at 16$\mu$m and one
of these is also not detected at 24$\mu$m.  Template fitting was not
performed for these three sources because there were not enough long
wavelength points to derive meaningful constraints. 

\section {Description of Individual Galaxy SEDs}

\subsection{Galaxies that show PAH emission}

\noindent \underline{KISSR 2344: } has 8$\mu$m emission that is significantly in
excess of the best-fitting SED template model. As can be seen in Table
\ref{tab:SED}, this galaxy also has one of the largest values for
f$_{\nu}$(8)/F$_{\nu}$(24), f$_{\nu}$(8)/F$_{\nu}$(16), and R$_1$. While this
galaxy probably does have some of the largest values for each of these
parameters, differences in the size of the apertures used at each of these
wavelengths could affect these numbers. A difference between the apertures would 
tend to increase the fluxes at the shorter wavelengths with respect to the longer 
wavelengths. This is the only galaxy in the sample for which this should be a 
significant issue since most of the others are unresolved or barely resolved.

\noindent \underline{KISSR 2406: } has a metallicity of $\log$[O/H]$+12 = 8.3$
and PAH emission in excess of the best-fit template models. In addition, this 
galaxy falls well within the PAH-emitting galaxy regions in Figures 
\ref{fig:engplt1} and \ref{fig:vsg_diag} and has a small fraction of its
infrared luminosity coming from the highest intensity regions (f$_{PDR} = 6-7$).

\noindent \underline{KISSR 2316: } is the highest metallicity system in the
sample ($\log$[O/H]$+12 = 8.8$) and has PAH emission that is only 1$\sigma$ 
below the best-fit model template. As with KISSR 2406, the galaxy falls well 
within the PAH-emitting 
galaxy regions in Figures \ref{fig:engplt1} and \ref{fig:vsg_diag} and has a
small fraction of its infrared luminosity coming from high intensity regions.

\noindent \underline{KISSR 2398: } has a metallicity of $\log$[O/H]$+12 = 8.4$
and 8$\mu$m emission that is only 1$\sigma$ below the best-fit model template.
The diagnostics from the other plots, Figures \ref{fig:engplt1}
and \ref{fig:vsg_diag}, also agreeing with this being a PAH emitting galaxy.

\noindent \underline{KISSR 2382: } has a metallicity of $\log$[O/H]$+12 = 8.2$
and 8$\mu$m emission that is 3$\sigma$ below the best-fit model template.
However, the galaxy falls well within the PAH-emitting galaxy regions in Figures 
\ref{fig:engplt1} and \ref{fig:vsg_diag}.

\subsection{Galaxies that show no evidence of PAH emission}

\noindent \underline{KISSR 2326: } shows no evidence for PAH emission -- the
data are 30$\sigma$ below the best-fit template at 8.0~$\mu$m in this 
$\log$[O/H]$+12 = 8.2$ galaxy. This galaxy
falls well within the no-PAH regions in Figures \ref{fig:engplt1} and
\ref{fig:vsg_diag}.

\noindent \underline{KISSR 2346: } shows no evidence for PAH emission as the
data are 58$\sigma$ below the best-fit template at 7$\mu$m in this 
low metallicity ($\log$[O/H]$+12 = 7.8$) galaxy. This galaxy
falls well within the no-PAH regions in Figures \ref{fig:engplt1} and
\ref{fig:vsg_diag}.

\noindent \underline{KISSR 2300: } shows no evidence for PAH emission. The
best-fit dust template is 70$\sigma$ above the 8.0
$\mu$m flux from this galaxy. There is no evidence for a rise in the spectrum
above the continuum level at these wavelengths for this galaxy that has
$\log$[O/H]$+12 = 7.9$. In addition, this galaxy clearly occupies the no-PAH
region in Figures \ref{fig:engplt1} and \ref{fig:vsg_diag}.

\noindent \underline{KISSR 2338: } shows no evidence for PAH emission in this 
$\log$[O/H]$+12 = 8.1$ metallicity galaxy. Even the best-fit dust template,
which is the template for an ultraluminous infrared galaxy (ULIRG), gives 
an extremely poor fit to these data ($\chi ^2 = 16.3$). This galaxy shows an
excess at 5.8, 8.0, and 16 $\mu$m that seems to indicate an unusually large 
contribution from hot dust and/or very small grains. The galaxy falls well
within the no-PAH regions in Figures \ref{fig:engplt1} and
\ref{fig:vsg_diag}.

\noindent \underline{KISSR 2349: } shows no evidence for PAH emission in this 
galaxy which has a metallicity of $\log$[O/H]$+12 = 8.1$.
Even the best-fit template gives a poor fit to the data ($\chi ^2 = 9.8$).
Similar to KISSR 2338, the best-fit template is for an ULIRG and there appears
to be an excess (although not quite as great as for KISSR 2338) of hot dust
and/or very small grain emission. This galaxy
falls well within the no-PAH regions in Figures \ref{fig:engplt1} and
\ref{fig:vsg_diag}.

\noindent \underline{KISSR 2368: }  shows no evidence for PAH emission in this 
$\log$[O/H]$+12 = 8.0$ galaxy. Even the best-fit dust template which, similar to
KISSR 2338 and 2349, is a ULIRG template, provides an extremely poor fit 
($\chi ^2 = 21.9$). As with the other two galaxies, there is some evidence for
an excess of hot dust and/or very small grains. This galaxy
falls well within the no-PAH regions in Figures \ref{fig:engplt1} and
\ref{fig:vsg_diag}.

\subsection{Galaxies for which the existence of PAH emission is uncertain}

\begin{figure}
\plotone{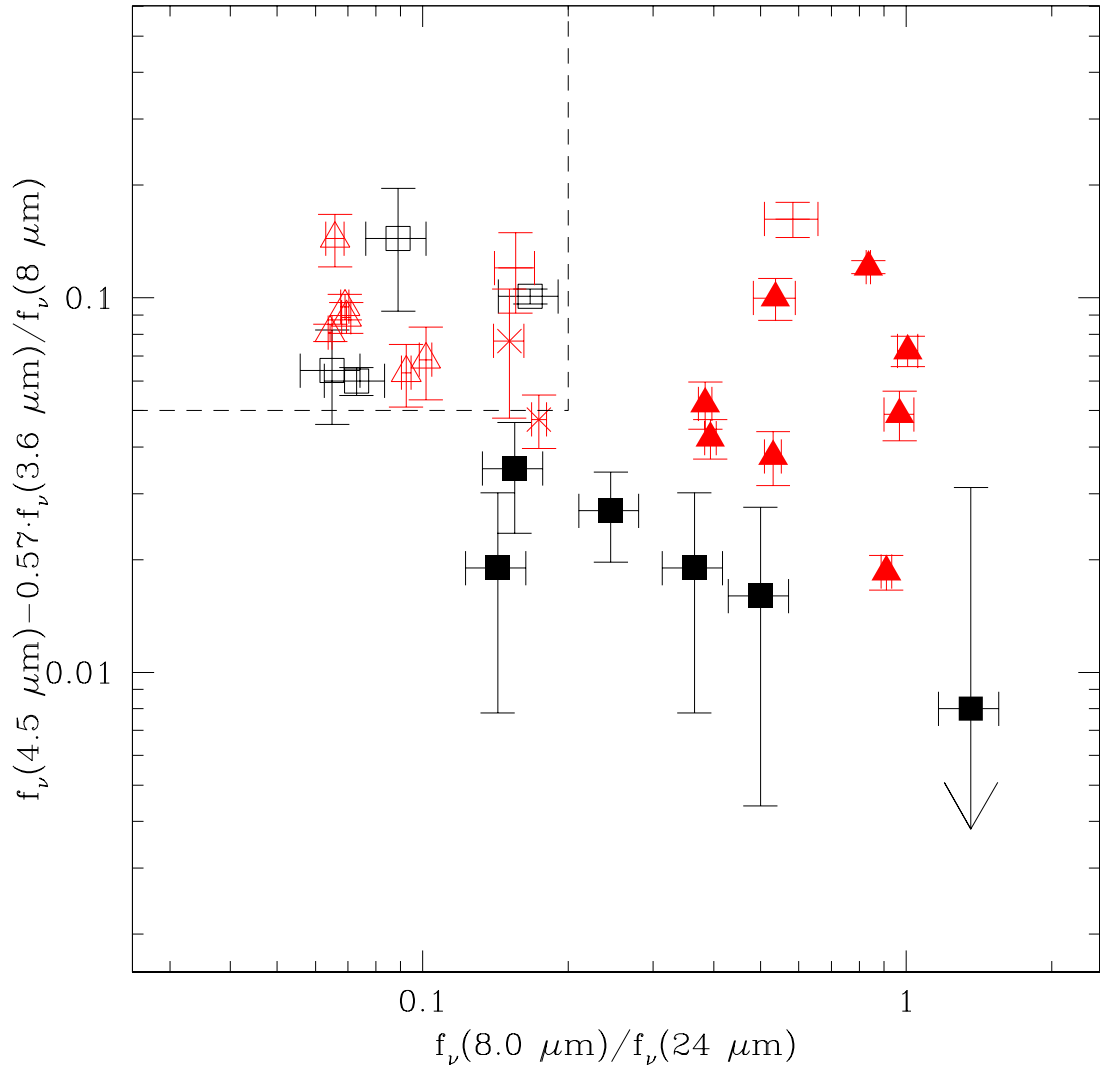}
\caption{The relationship between two different MIR flux ratios for the galaxies. 
The 8 $\mu$m flux includes PAH, dust continuum, and stellar emission, but the
stellar emission is very small and would not introduce a significant shift in
these data. The y-axis ratio is the stellar subtracted 4.5$\mu$m continuum to 
the 8 $\mu$m flux. The region enclosed by the dashed lines is the ``no-PAH" 
region as defined from the \citet{engelbracht2005} sample. The black
points are from \citet{engelbracht2005} -- the open squares are
spectroscopically confirmed to lack PAH emission, the filled squares are
spectroscopically confirmed to possess PAH emission. The red points show the
galaxies from our sample. Even though none of them have {\it Spitzer}/IRS
spectroscopy to determine the presence of PAH emission we mark in filled/open
triangles the ones for which the SEDs indicate the presence/absence of PAH
emission. The galaxies for which the presence or absence of PAH emission could
not be determined are marked with $\times$s. For the galaxies that do not have 
enough data to construct SEDs, only errorbars are plotted.}
\label{fig:engplt1}
\end{figure}

\begin{figure}
\plotone{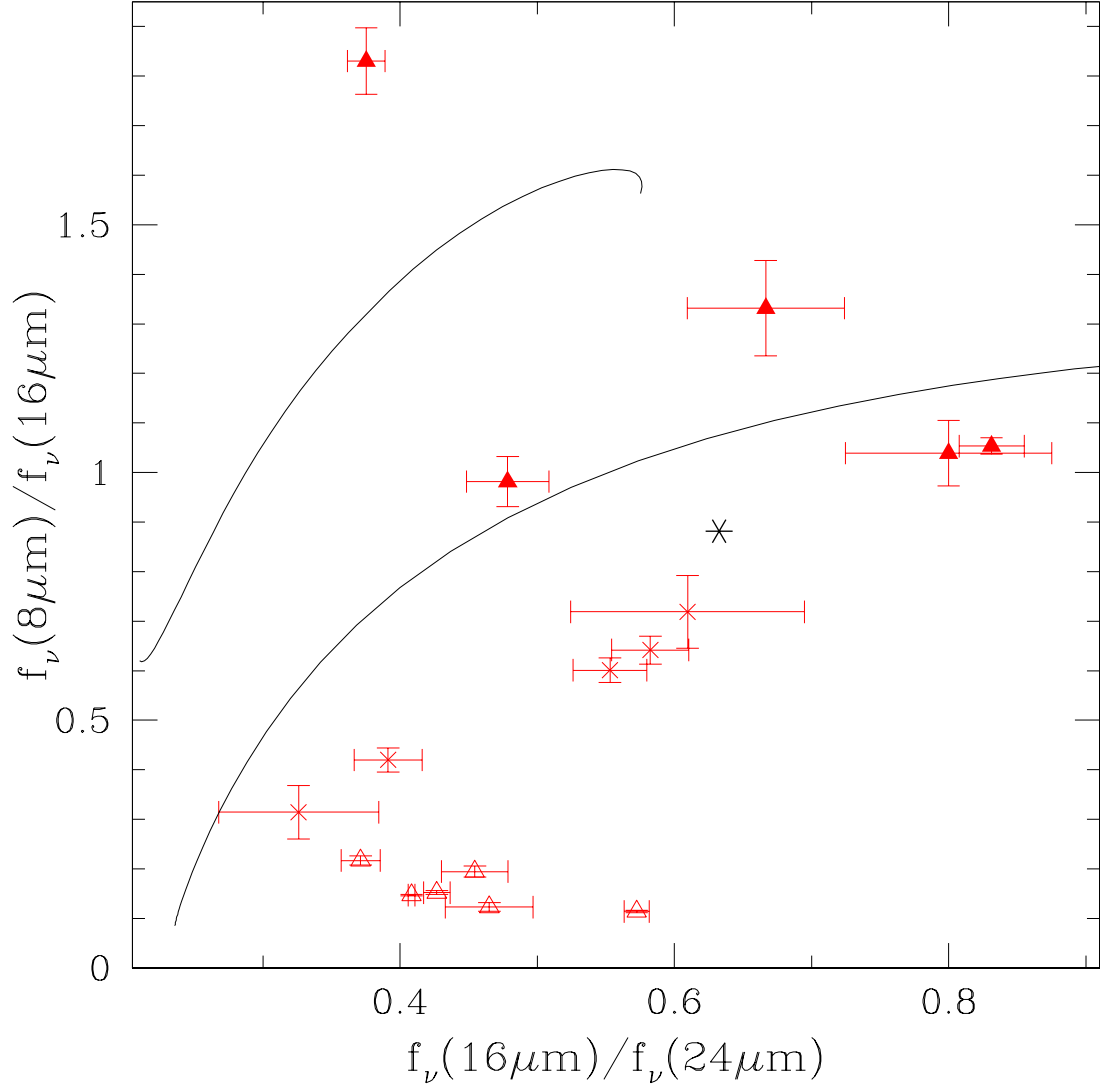}
\caption{The mid-infrared color ratio for galaxies in our sample (red
points). The filled symbols represent galaxies with PAH emission while the open
symbols represent galaxies without PA emission. Even though none of
these galaxies have {\it Spitzer}/IRS spectroscopy to determine the presence of
PAH emission we mark in filled/open triangles as determined by the SED fitting.
The $\times$s are galaxies for which we were not able to determine the presence
or absence of PAH emission from the SEDs. The solid lines are the
tracks followed by star-forming galaxies from the \citet{dale2002} (lower line)
and \citet{lagache2003} (upper line) samples, all of which exhibit PAH
emission. The higher luminosity galaxies are at the bottom right end of the
tracks. The black star is the ``normal" galaxy from \citet{lagache2003}.}
\label{fig:vsg_diag}
\end{figure}

\begin{figure}
\plotone{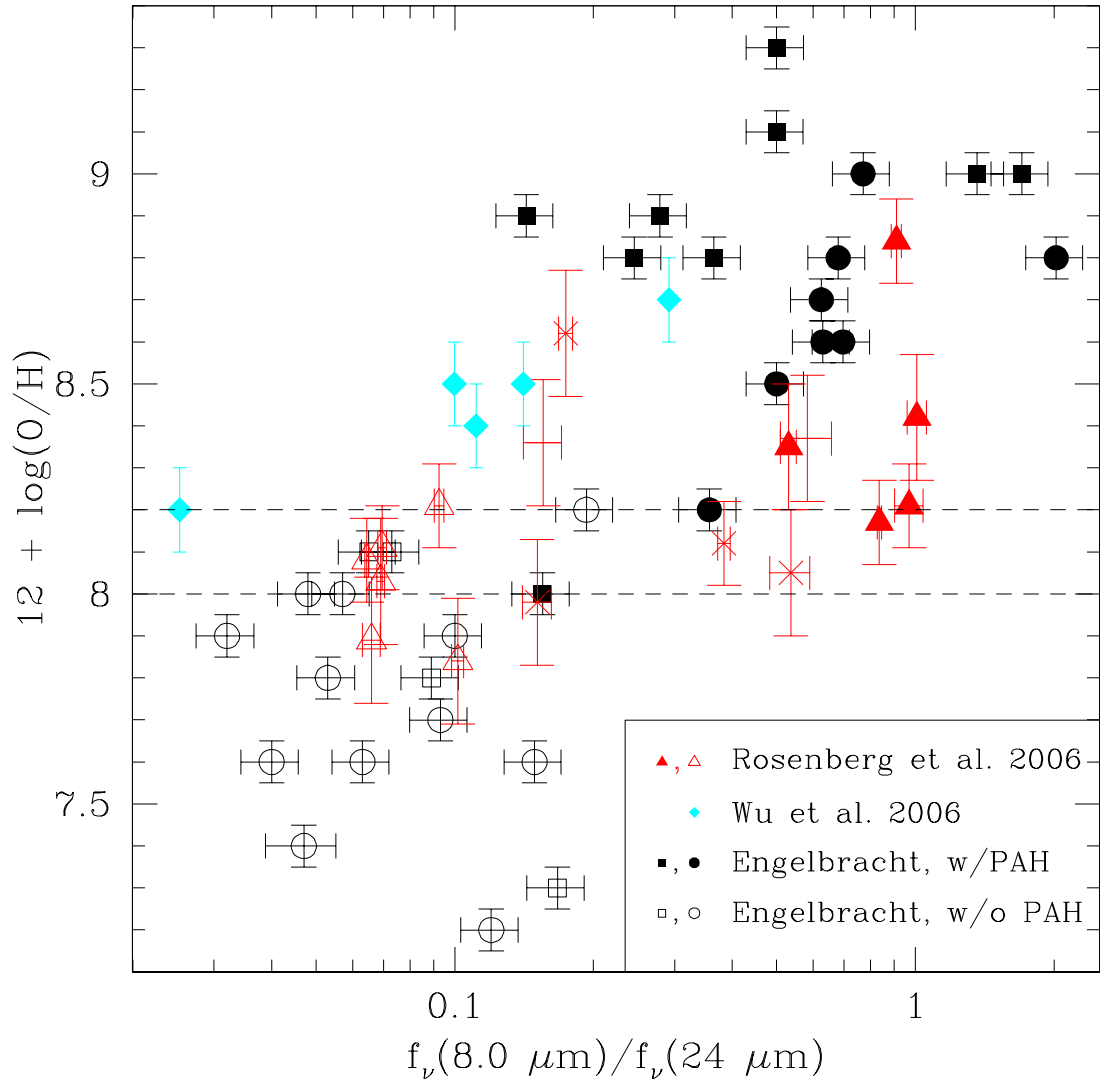}
\caption{Galaxy metallicity as a function of the 8 to 24 $\mu$m flux density
ratio. The
black points are galaxies from \citet{engelbracht2005}, the blue diamonds are
blue compact dwarfs from \citet{wu2006}, and the red points are star-forming
dwarf galaxies in this sample. Open points are systems that lack PAH emission 
(determined from either spectra or Figure \ref{fig:engplt1} for the galaxies
from \citet{engelbracht2005}, from spectra for the galaxies from \citet{wu2006},
and from the broadband SEDs for the galaxies in this sample) while the solid
points are thought to exhibit PAH emission. Errorbars with no points are
galaxies in our sample that were not detected at the longer wavelengths so SEDs
could not be constructed. The dashed lines delineate the region between
$\log[O/H]+12=8.0$ and 8.2}.
\label{fig:abundrat}
\end{figure}

\noindent \underline{KISSR 2309: } is a low metallicity galaxy ($\log$[O/H]$+12
= 8.0$) which has an 8~$\mu$m flux density that is 7$\sigma$ below the best-fit
SED template. In Figure \ref{fig:engplt1} the galaxy falls within the no-PAH
region. In Figure \ref{fig:vsg_diag},
KISSR 2309 is the left-most galaxy in the no-PAH region with errorbars that
touch the tracks for the spectral templates which include PAH emission. 

\noindent \underline{KISSR 2322: } is another low metallicity galaxy
($\log$[O/H]$+12 = 8.0$) which has an 8~$\mu$m flux density that is 7$\sigma$
below the best-fit SED template. In contrast to KISSR 2309, KISSR 2322 falls
within the PAH-emitting region in Figure \ref{fig:engplt1} and between the PAH
and no-PAH galaxies in Figure \ref{fig:vsg_diag}.

\noindent \underline{KISSR 2359: } is a high metallicity galaxy ($\log$[O/H]$+12
= 8.6$) yet it has an 8~$\mu$m flux density that is 14$\sigma$ below the
best-fit SED template. In Figure \ref{fig:engplt1} the
galaxy falls just outside the no-PAH region, although the errorbars overlap with
the region. In Figure \ref{fig:vsg_diag} KISSR 2359 has the second lowest value
of f$_{\nu}$(8)/f$_{\nu}$(16) of the galaxies for which the PAH classification
is uncertain, placing it closer to the no-PAH galaxies than the PAH emitting
ones.

\noindent \underline{KISSR 2318: } has an unknown metallicity and an 8~$\mu$m
flux density that is 12$\sigma$ below the best-fit SED template. Nevertheless,
this galaxy falls well within the PAH-emitting galaxy region in Figures 
\ref{fig:engplt1} and has one of the higher f$_{\nu}$(8)/f$_{\nu}$(16) ratios of
the galaxies with uncertain PAH emission in Figure \ref{fig:vsg_diag}.

\noindent \underline{KISSR 2292:} is another fairly low metallicity galaxy
($\log$[O/H]$+12 = 8.1$) which has an 8~$\mu$m flux density that is 13$\sigma$
below the best-fit SED template. The galaxy falls in the PAH-emitting region in
Figure \ref{fig:engplt1} and has one of the higher f$_{\nu}$(8)/f$_{\nu}$(16)
ratios of the galaxies with uncertain PAH emission in Figure \ref{fig:vsg_diag}.

%\bibliographystyle{astron}
%\bibliography{references}

\clearpage
\begin{deluxetable}{cccrcccc}
\tabletypesize{\small}
\tablewidth{5.8in}
\tablecaption{Properties of KISS Galaxies \label{tab:STATS}}
\tablehead{
\colhead{KISSR} & \colhead{RA} & \colhead{Dec} & 
\colhead{v$_{hel}$\tablenotemark{a}} & \colhead{M$_{B0}$\tablenotemark{b}} & 
\colhead{c$_{H\beta}$\tablenotemark{c}} & 
\colhead{$\log L_{\rm H\alpha}$} & \colhead{$\log[O/H]+12$} \\
\colhead{} & \colhead{J2000} & \colhead{J2000} & 
\colhead{km s$^{-1}$} & \colhead{} & \colhead{} & 
\colhead{ergs s$^{-1}$} & \colhead{}}
\startdata
 2292 & 14:25:09.2  & 35:25:15.9  &  8659 & -18.31 &  0.283 & 40.39 &  8.12 \\
 2300 & 14:26:08.9  & 33:54:19.8  & 10271 & -16.09 &  0.154 & 40.52 &  7.89 \\
 2302 & 14:26:17.5  & 35:21:35.5  &  8342 & -17.16 & \nodata &  40.04 & \nodata \\
 2309 & 14:26:53.6  & 34:04:14.5  &  7231 & -17.12 &  0.284 & 40.09 &  7.98 \\
 2316 & 14:28:14.9  & 33:30:25.7  & 10685 & -18.57 &  1.094 & 41.01 &  8.84 \\
 2318 & 14:28:24.6  & 35:10:21.5  & 22163 & -17.88 & \nodata &  41.15 & \nodata \\
 2322 & 14:29:09.6  & 32:51:26.7  &  8574 & -17.84 &  0.096 & 40.24 &  8.05 \\
 2326 & 14:29:32.7  & 33:30:40.3  &  7935 & -17.14 &  0.151 & 40.65 &  8.21 \\
 2338 & 14:30:27.9  & 35:32:07.2  & 11689 & -17.20 &  0.190 & 40.81 &  8.11 \\
 2344 & 14:31:03.6  & 35:31:14.8  &  4166 & -17.75 &  0.248 & 40.12 &  8.17 \\
 2346 & 14:31:14.4  & 33:19:13.2  & 10819 & -16.80 &  0.197 & 40.49 &  7.84 \\
 2349 & 14:31:20.0  & 34:38:03.8  &  4396 & -16.63 & -0.012 & 40.48 &  8.08 \\
 2357 & 14:31:39.2  & 33:26:32.3  & 10759 & -17.67 &  0.504 & 40.40 &  8.37 \\
 2359 & 14:31:49.3  & 35:28:40.0  & 22512 & -18.03 &  1.350 & 41.66 &  8.62 \\
 2368 & 14:32:18.9  & 33:02:53.7  & 10972 & -17.05 &  0.193 & 40.84 &  8.03 \\
 2382 & 14:34:08.0  & 34:19:34.4  &  6813 & -17.81 &  0.100 & 40.03 &  8.21 \\
 2398 & 14:36:33.1  & 34:58:04.4  &  9006 & -18.01 &  0.187 & 40.33 &  8.42 \\
 2403 & 14:37:42.6  & 33:36:26.7  & 12047 & -16.24 &  0.362 & 40.38 &  8.36 \\
 2406 & 14:38:27.8  & 35:08:59.0  &  8641 & -17.89 &  0.185 & 40.38 &  8.35 \\
\enddata
 \tablenotetext{a}{Heliocentric velocity}
 \tablenotetext{b}{B-band absolute magnitude corrected for Galactic extinction}
 \tablenotetext{c}{Optical reddening at H$\beta$}
\end{deluxetable}

%\clearpage
\begin{deluxetable}{crcrcrrcccccc}
\tabletypesize{\small}
\tablewidth{5.8in}
\tablecaption{Infrared and Ultraviolet Fluxes of KISS Galaxies \label{tab:FLUX}}
\tablehead{
\colhead{KISSR} & 
\colhead{f$_{16}$} & \colhead{$\sigma_{16}$} &
\colhead{f$_{24}$} & \colhead{$\sigma_{24}$} &
\colhead{f$_{70}$} & \colhead{$\sigma_{70}$} &
\colhead{f$_{160}$} & \colhead{$\sigma_{160}$} &
\colhead{f$_{NUV}$} & \colhead{$\sigma_{NUV}$} &
\colhead{f$_{FUV}$} & \colhead{$\sigma_{FUV}$} \\
\colhead{} & \colhead{mJy} &
\colhead{mJy} & \colhead{mJy} & \colhead{mJy} & \colhead{mJy} &
\colhead{mJy} & \colhead{mJy} & \colhead{mJy} & \colhead{$\mu$Jy} &
\colhead{$\mu$Jy} & \colhead{$\mu$Jy} & \colhead{$\mu$Jy}}
\startdata
2292 &    1.27  & 0.08    &   2.35    & 0.07    &  \nodata &  \nodata &  $<$120  &  \nodata &   74.98 & 0.69  & \nodata & \nodata \\ 
2300 &    0.77  & 0.05    &   1.72    & 0.05    &  $<$15   &  \nodata &  $<$120  &  \nodata &   31.26 & 0.50  & \nodata & \nodata \\ 
2302 &  $<$0.1 & \nodata & $<$0.24   & \nodata &  $<$15   &  \nodata &  $<$120  &  \nodata &   92.42 & 0.84  & \nodata & \nodata \\
2309 &    0.32  & 0.02    &   0.92    & 0.06    &  $<$15   &  \nodata &  $<$120  &  \nodata &  111.00 & 0.88  & \nodata & \nodata \\
2316 &    3.39  & 0.20    &   4.09    & 0.10    &   52	  &    4     &   150    &      33  &   19.40 & 0.33  & \nodata & \nodata \\
2318 &    1.17  & 0.07    &   2.06    & 0.05    &   15	  &    4     &  $<$120  &  \nodata &    7.96 & 0.29  & \nodata & \nodata \\
2322 &    0.49  & 0.03    &   0.82    & 0.08    &  $<$15   &  \nodata &  $<$100  &  \nodata &  154.30 & 1.31  & \nodata & \nodata \\
2326 &    1.43  & 0.08    &   3.77    & 0.05    &   29	  &    3     &  $<$100  &  \nodata &  117.80 & 0.80  & \nodata & \nodata \\
2338 &    2.85  & 0.17    &   6.56    & 0.09    &   25	  &    5     &  $<$100  &  \nodata &   49.46 & 0.55  &  39.20  &  0.81 \\
2344 &    1.41  & 0.08    &   3.73    & 0.03    &   71	  &    6     &   210    &	33 &  581.44 & 2.04  & 439.02  &  2.47 \\
2346 &    0.98  & 0.06    &   2.20    & 0.04    &  $<$15   &  \nodata &  $<$120  &  \nodata &   27.23 & 0.54  & \nodata & \nodata \\
2349 &   10.72  & 0.64    &  26.20    & 0.09    &  154	  &    7     &   190    &	33 &  276.85 & 1.53  & \nodata & \nodata \\
2357 &  $<$0.3 & \nodata &  0.48     & 0.06    &  $<$15   &  \nodata &  $<$120  &  \nodata &   75.33 & 0.95  & \nodata & \nodata \\
2359 &    0.94  & 0.05    &   2.30    & 0.07    &   20	  &    4     &  $<$100  &  \nodata &   11.87 & 0.31  &   7.88  &  0.34 \\
2368 &    4.34  & 0.26    &   7.51    & 0.08    &   30	  &   10     &  $<$120  &  \nodata &  113.90 & 1.24  & \nodata & \nodata \\
2382 &    0.81  & 0.05    &   1.00    & 0.07    &   34	  &    7     &  $<$120  &  \nodata &  242.74 & 1.40  & 140.28  &  5.80 \\
2398 &    0.74  & 0.04    &   1.05    & 0.05    &   20	  &    3     &  $<$120  &  \nodata &  128.21 & 0.36  &  92.40  &  0.44 \cr
2403 &  $<$0.2 & \nodata &  0.47     & 0.04    &  $<$15   &  \nodata & \nodata  &  \nodata &   21.20 & 0.47  & \nodata & \nodata \cr
2406 &    1.00  & 0.06    &   2.09    & 0.08    &   27	  &    6     &  $<$120  &  \nodata &  164.71 & 0.44  & 126.32  &  0.50 \cr
\enddata
\end{deluxetable}
\clearpage

\clearpage
\begin{landscape}
\begin{deluxetable}{lccrcrrcccrc}
%\tabletypesize{\small}
\tablewidth{8.1in}
\tablecaption{Properties Influenced by PAH Emission in KISS Galaxies \label{tab:SED}}
\tablehead{
\colhead{KISSR} & 
\colhead{f[8]$_{model}$\tablenotemark{a}} &
\colhead{f[8]$_{data}$\tablenotemark{b}} & 
\colhead{f[8]$_{diff}$\tablenotemark{c}} & \colhead{$\sigma$[8]\tablenotemark{d}} &
\colhead{$\sigma_{diff}$\tablenotemark{e}} &
\colhead{$\chi_{red}^2$\tablenotemark{f}} & 
\colhead{f[8]/f[24]\tablenotemark{g}} & \colhead{f[8]/f[16]\tablenotemark{h}} & 
\colhead{R$_1$\tablenotemark{i}} &
\colhead{f$_{PDR}$\tablenotemark{j}} &
\colhead{PAH?\tablenotemark{k}} \\
\colhead{} & \colhead{mJy} & \colhead{mJy} & \colhead{mJy} & \colhead{mJy} &
\colhead{} & \colhead{} & \colhead{} & \colhead{} & \colhead{} & \colhead{\%}}
\startdata
2344 & 1.99 & 3.07 & -1.08 & 0.090 & 12 &  3.2 & 0.84 & 1.83 & 0.12 & 4       & Y \\
2406 & 0.96 & 1.08 & -0.12 & 0.034 &  4 &  1.0 & 0.53 & 0.98 & 0.04 & 6-7     & Y \\
2316 & 3.49 & 3.61 & -0.12 & 0.109 &  1 &  3.1 & 0.91 & 1.05 & 0.02 & 8       & Y \\
2398 & 1.04 & 1.03 &  0.01 & 0.033 &  1 &  0.8 & 1.01 & 1.33 & 0.07 & 0       & Y \\
2382 & 1.04 & 0.95 &  0.09 & 0.030 &  3 &  3.2 & 0.97 & 1.04 & 0.05 & 0       & Y \\
\hline
2309 & 0.17 & 0.14 &  0.03 & 0.006 &  7 &  1.3 & 0.15 & 0.31 & 0.08 & \nodata & ? \\
2322 & 0.54 & 0.43 &  0.11 & 0.015 &  7 &  0.04 & 0.54 & 0.72 & 0.10 & \nodata & ? \\
2359 & 0.55 & 0.37 &  0.18 & 0.013 & 14 &  0.4 & 0.17 & 0.42 & 0.05 & 13      & ? \\
2318 & 1.03 & 0.75 &  0.28 & 0.024 & 12 &  0.5 & 0.39 & 0.64 & 0.04 & \nodata & ? \\
2292 & 1.24 & 0.88 &  0.36 & 0.028 & 13 &  1.4 & 0.38 & 0.60 & 0.05 & \nodata & ? \\ 
\hline
2326 & 0.67 & 0.37 &  0.30 & 0.010 & 30 &  2.3 & 0.09 & 0.22 & 0.06 & 15-20   & N \\
2346 & 0.68 & 0.22 &  0.46 & 0.008 & 58 &  2.9 & 0.10 & 0.19 & 0.07 & \nodata & N \\
2300 & 0.46 & 0.11 &  0.35 & 0.005 & 70 &  3.1 & 0.07 & 0.12 & 0.14 & \nodata & N \\ 
\hline
2338 & 0.17 & 0.44 & -0.27 & 0.015 & 18 & 16.3 & 0.07 & 0.15 & 0.09 & 31-35   & N \\
2349 & 1.74 & 1.66 &  0.08 & 0.050 &  2 &  9.8 & 0.06 & 0.15 & 0.08 & 40      & N \\
2368 & 0.94 & 0.50 &  0.44 & 0.017 & 26 & 21.9 & 0.07 & 0.11 & 0.09 & 30-34   & N \\
\hline
2302&\nodata&      &\nodata&  &\nodata&\nodata&\nodata&\nodata&\nodata&\nodata&\nodata \\
2357&\nodata&      &\nodata&  &\nodata&\nodata&0.58&\nodata&0.16&\nodata&\nodata \\
2403&\nodata&      &\nodata&  &\nodata&\nodata&0.16&\nodata&0.12&\nodata&\nodata
\enddata
\tablenotetext{a}{The 8$\mu$m flux of the best-fit template SED}
\tablenotetext{b}{The 8$\mu$m flux measured for the galaxy}
\tablenotetext{c}{The difference between the model and the galaxy 8$\mu$m flux}
\tablenotetext{d}{The error in the measured 8$\mu$m flux}
\tablenotetext{e}{The number of $\sigma$ difference between the model and the
data at 8$\mu$m}
\tablenotetext{f}{The reduced $\chi^2$ of the fit between the data and the
template SED}
\tablenotetext{g}{The 8 to 24 $\mu$m flux ratio}
\tablenotetext{h}{The 8 to 16 $\mu$m flux ratio}
\tablenotetext{i}{R$_1 =$
(f$_{\nu}$(4.5$\mu$m)-0.57f$_{\nu}$(3.6$\mu$m))/f$_{\nu}$(8$\mu$m)}
\tablenotetext{j}{The fraction of the infrared emission coming from regions with
U$ > 10^2$ in the \citet{draine2007} models}
\tablenotetext{k}{The assessment of whether the galaxies has significant PAH
emission}
\end{deluxetable}
\clearpage
\end{landscape}

\clearpage
\begin{landscape}
\begin{deluxetable}{crccrrcccr}
\tablewidth{6.8in}
\tabletypesize{\footnotesize}
\tablecaption{Infrared, Optical, and UV SFRs of KISS Galaxies \label{tab:SFR}}
\tablehead{
\colhead{KISSR} & \colhead{SFR$_{CE}$\tablenotemark{a,b}} & \colhead{SFR$_{8\mu
m}$\tablenotemark{a,c}} & \colhead{SFR$_{24\mu m}$\tablenotemark{a,d}} &
\colhead{SFR$_{Calz}$\tablenotemark{a,e}} &
\colhead{SFR$_{H\alpha}$\tablenotemark{a,f}} &
\colhead{SFR$_{H\alpha _Z}$\tablenotemark{a,g}} &
\colhead{SFR$_{TIR}$\tablenotemark{a,h}} & 
\colhead{SFR$_{NUV}$\tablenotemark{a,i}} &
\colhead{SFR$_{FUV}$\tablenotemark{a,j}} }
\startdata
 2292 &  0.30  &  0.10 &    0.21 &  0.09  &  0.19 &  0.07 & \nodata & 0.19 & \nodata \\
 2300 &  0.32  &  0.02 &    0.22 &  0.02  &  0.26 &  0.08 & \nodata & 0.11 & \nodata \\
 2302 & $<$0.05 & 0.01 & $<$0.02 & \nodata&  0.09 &  0.09 & \nodata & 0.22 & \nodata \\
 2309 &  0.09  &  0.01 &    0.06 &  0.01  &  0.10 &  0.03 & \nodata & 0.20 & \nodata \\
 2316 &  0.70  &  0.65 &    0.56 &  0.48  &  0.81 &  0.68 & 0.79    & 0.08 & \nodata \\
 2318 &  1.66  &  0.61 &    1.21 &  0.54  &  1.12 &  1.12 & 0.5-2.1 & 0.13 & \nodata \\
 2322 &  0.12  &  0.05 &    0.07 &  0.04  &  0.14 &  0.05 & \nodata & 0.39 & \nodata \\
 2326 &  0.38  &  0.03 &    0.28 &  0.04  &  0.35 &  0.15 & 0.1-0.3 & 0.25 & \nodata \\
 2338 &  1.23  &  0.09 &    1.07 &  0.12  &  0.51 &  0.19 & 0.3-0.7 & 0.23 &  0.18   \\
 2344 &  0.12  &  0.08 &    0.08 &  0.06  &  0.10 &  0.04 & 0.16    & 0.34 &  0.26   \\
 2346 &  0.41  &  0.04 &    0.31 &  0.05  &  0.24 &  0.07 & \nodata & 0.11 & \nodata \\
 2349 &  0.64  &  0.05 &    0.61 &  0.06  &  0.24 &  0.09 & 0.32    & 0.18 & \nodata \\
 2357 &  0.11  &  0.05 &    0.07 &  0.04  &  0.20 &  0.09 & \nodata & 0.30 & \nodata \\
 2359 &  1.87  &  0.31 &    1.40 &  0.32  &  3.59 &  2.09 & 0.6-2.0 & 0.21 &  0.14   \\
 2368 &  1.26  &  0.09 &    1.08 &  0.12  &  0.54 &  0.18 & 0.3-1.0 & 0.47 & \nodata \\
 2382 &  0.09  &  0.07 &    0.06 &  0.05  &  0.08 &  0.04 & 0.1-0.2 & 0.38 &  0.22   \\
 2398 &  0.15  &  0.13 &    0.10 &  0.09  &  0.17 &  0.08 & 0.1-0.4 & 0.36 &  0.26   \\
 2403 &  0.13  &  0.02 &    0.08 &  0.02  &  0.19 &  0.09 & \nodata & 0.10 & \nodata \\
 2406 &  0.27  &  0.13 &    0.19 &  0.10  &  0.19 &  0.09 & 0.1-0.4 & 0.42 &   0.32  \\ 
\enddata
\tablenotetext{a}{The units are M\solar\ yr$^{-1}$}
\tablenotetext{b}{SFR from the \citet{chary2001} models.}
\tablenotetext{c}{SFR from the 8~$\mu$m \citet{hwu2005} relation.}
\tablenotetext{d}{SFR from the 24~$\mu$m \citet{hwu2005} relation.}
\tablenotetext{e}{SFR from a combination of 8~$\mu$m and 24~$\mu$m
flux as described by \citet{calzetti2005}}
\tablenotetext{f}{\ha\ SFR from the \citet{kennicutt1998} relation.}
\tablenotetext{g}{\ha\ SFR from the \citet{lee2002} relation which
accounts for the harder radiation field in low metallicity galaxies.}
\tablenotetext{h}{SFR from the 24, 70, and 160~$\mu$m fluxes using the \citet{dale2005} and
\citet{kennicutt1998} relations.}
\tablenotetext{i}{SFR calculated from the NUV flux using the
\citet{kennicutt1998} relation.}
\tablenotetext{j}{SFR calculated from the FUV flux using the
\citet{kennicutt1998} relation.}
\end{deluxetable}
\clearpage
\end{landscape}

\end{document}